\documentclass[usegraphicx,usenatbib]{mn2e}

\usepackage{fixltx2e}
\usepackage{mathptmx}
\usepackage{amsmath}
\usepackage{url}
\usepackage{times}

\newcommand{\Mpc}{\rm\; Mpc}
\newcommand{\kpc}{\rm\; kpc}

\newcommand{\km}{\rm\; km}

\newcommand{\cm}{\rm\; cm}

\newcommand{\yr}{\rm\; yr}
\newcommand{\Gyr}{\rm\; Gyr}

\newcommand{\s}{\rm\; s}
\newcommand{\ks}{\rm\; ks}

\newcommand{\GHz}{\rm\; GHz}
\newcommand{\MHz}{\rm\; MHz}
\newcommand{\Hz}{\rm\; Hz}
\newcommand{\Msun}{\hbox{$\rm\thinspace M_{\odot}$}}

\newcommand{\Msunpyr}{\hbox{$\Msun\yr^{-1}\,$}}
\newcommand{\keV}{\rm\; keV}

\newcommand{\erg}{\rm\; erg}

\newcommand{\mJy}{\rm\; mJy}
\newcommand{\W}{\rm\; W}

\newcommand{\ergps}{\hbox{$\erg\s^{-1}\,$}}

\newcommand{\WpHz}{\hbox{$\W\pHz\,$}}
\newcommand{\surbri}{\hbox{$\rm\thinspace photons\pcmsq\ps\pasecsq$}}
\newcommand{\kmps}{\hbox{$\km\s^{-1}\,$}}

\newcommand{\kmpspMpc}{\hbox{$\kmps\Mpc^{-1}\,$}}
\newcommand{\Zsun}{\hbox{$\thinspace \mathrm{Z}_{\odot}$}}

\newcommand{\chisq}{\hbox{$\chi^2$}}

\newcommand{\amin}{\rm\; arcmin}

\newcommand{\asec}{\rm\; arcsec}

\newcommand{\emm}{\hbox{$\cm^{-5}\,$}}
\newcommand{\empasecsq}{\hbox{$\emm\asec^{-2}\,$}}
\newcommand{\keVempasecsq}{\hbox{$\keV\emm\asec^{-2}\,$}}
\newcommand{\pcmsq}{\hbox{$\cm^{-2}\,$}}
\newcommand{\pcmcu}{\hbox{$\cm^{-3}\,$}}

\newcommand{\ps}{\hbox{$\s^{-1}\,$}}
\newcommand{\pHz}{\hbox{$\Hz^{-1}\,$}}

\newcommand{\pasecsq}{\hbox{$\asec^{-2}\,$}}
\begin{document}

\title[The merging cluster Abell 2146]{\emph{Chandra} observation of
  two shock fronts in the merging galaxy cluster Abell 2146}
\author[H.R. Russell et al.]
%{\parbox[]{7.in}{H.~R. Russell\thanks{E-mail:
%      hrr27@ast.cam.ac.uk}, et al.}
{\parbox[]{7.in}{H.~R. Russell$^1$\thanks{E-mail: hrr27@ast.cam.ac.uk},
    J.~S. Sanders$^1$, A.~C. Fabian$^1$, S.~A. Baum$^2$, M. Donahue$^3$, A.~C. Edge$^4$,
    B.~R. McNamara$^5$ and C.~P. O'Dea$^6$ \\
    \footnotesize
    $^1$ Institute of Astronomy, Madingley Road, Cambridge CB3 0HA\\
    $^2$ Center for Imaging Science, Rochester Institute of
    Technology, Rochester, NY 14623, USA\\
    $^3$ Department of Physics and Astronomy, Michigan State University,
    East Lansing, MI 48824, USA\\
    $^4$ Department of Physics, Durham University, Durham DH1 3LE\\
    $^5$ Department of Physics and Astronomy, University of Waterloo,
    Waterloo, ON N2L 3G1, Canada\\
    $^6$ Department of Physics, Rochester Institute of Technology,
    Rochester, NY 14623, USA\\
  }
}
\maketitle

\begin{abstract}
  We present a new \emph{Chandra} observation of the galaxy cluster
  Abell 2146 which has revealed a complex merging system with a gas
  structure that is remarkably similar to the Bullet cluster
  (eg. \citealt{Markevitch02}).  The X-ray image and temperature map
  show a cool $2-3\keV$ subcluster with a ram pressure stripped tail of gas
  just exiting the disrupted $6-7\keV$ primary cluster.  From the
  sharp jump in the temperature and density of the gas, we determine
  that the subcluster is preceded by a bow shock with a Mach number
  $M=2.2\pm0.8$, corresponding to a velocity
  $v=2200^{+1000}_{-900}\kmps$ relative to the main cluster.  We
  estimate that the subcluster passed through the primary core only
  $0.1-0.3\Gyr$ ago.  In addition, we observe a slower upstream shock
  propagating through the outer region of the primary cluster and
  calculate a Mach number $M=1.7\pm0.3$.  Based on the measured shock
  Mach numbers $M\sim2$ and the strength of the upstream shock, we
  argue that the mass ratio between the two merging clusters is
  between 3 and 4 to one.  By comparing the \emph{Chandra} observation with an
  archival HST observation, we find that a group of galaxies is
  located in front of the X-ray subcluster core but the brightest
  cluster galaxy is located immediately behind the X-ray peak.
\end{abstract}


\section{Introduction}
% Paragraph on cluster mergers
Galaxy clusters are assembled by hierarchical mergers of smaller
subclusters and groups.  These subclusters collide at velocities of a
few thousand $\kmps$, releasing as much as $10^{64}\erg$ of kinetic
energy as thermal energy by driving shocks, generating turbulence and
probably accelerating relativistic particles (see
eg. \citealt{Sarazin01}; \citealt{Feretti02}).  Major cluster mergers
are therefore the most energetic events since the Big Bang.  Shock
fronts provide a key observational tool in the study of these systems.
They can be used to determine the velocity and kinematics of the
merger and to study the conditions and transport processes in the ICM,
including electron-ion equilibrium and thermal conduction
(eg. \citealt{Markevitch06}).  Combining X-ray observations of merging
clusters with gravitational lensing studies has also produced
direct detections of dark matter (\citealt{Clowe04,Clowe06};
\citealt{Bradac06}) and constraints on the dark matter
self-interaction cross section (\citealt{Markevitch04};
\citealt{Randall08}).  Radio observations of supernova
remnants indicate that a fraction of the shock energy can be converted
into the acceleration of relativistic particles
(eg. \citealt{Blandford87}).  It is likely that this process also
operates in cluster mergers and could produce synchrotron radio
emission (eg. \citealt{Feretti02,Feretti05}; \citealt{Buote01};
\citealt{Kempner04}) and inverse Compton hard X-ray emission
(eg. \citealt{Fusco-Femiano99}; \citealt{Rephaeli99};
\citealt{Fusco-Femiano05}).  However, as radio emitting electrons have
short radiative lifetimes ($10^7-10^8\yr$), it is difficult to explain
the $\sim\Mpc$ size of extended radio halos (for a review see
eg. \citealt{Brunetti03}).  

X-ray observations of merging shocks currently provide the only method
for determining the velocity of the cluster gas in the plane of the sky
(eg. \citealt{Markevitch99}).  By measuring the temperature and
density of the gas on either side of the shock using X-ray imaging
spectroscopy, the shock velocity can be calculated from the
Rankine-Hugoniot jump
conditions.  While many clusters are found to have shock-heated
regions (eg. \citealt{Henry95,Henry96}; \citealt{Markevitch96};
\citealt{Belsole04,Belsole05}), the detection of a sharp density edge
and an unambiguous jump in temperature is rare.  Currently only two
shock fronts have been found by \emph{Chandra}: one in the Bullet
cluster (1E\,0657-56; \citealt{Markevitch02}; \citealt{Markevitch06})
and the other in Abell 520 (\citealt{Markevitch05}).  In this paper we
present two new merger shock fronts discovered in a recent
\emph{Chandra} observation of the galaxy cluster Abell 2146 at a
redshift $z=0.234$ (\citealt{Struble99}; \citealt{Bohringer00}).

We assume $H_0=70\kmpspMpc$, $\Omega_m=0.3$ and $\Omega_\Lambda=0.7$,
translating to a scale of $3.7\kpc$ per arcsec at the redshift
$z=0.234$ of Abell 2146.  All errors are $1\sigma$ unless otherwise
noted.

\section{Data Preparation}
% Chandra data reduction
Abell 2146 was observed with the \emph{Chandra} ACIS-S detector for
$43\ks$ split into two observations which were taken only a day apart
in April 2009 (Obs. IDs 10464 and 10888).  The data were analysed with
CIAO version 4.2 and CALDB version 4.2.0 provided by the
\emph{Chandra} X-ray Center (CXC).  The level 1 event files were
reprocessed to apply the latest gain and charge transfer inefficiency
correction and then filtered to remove photons detected with bad
grades.  The improved background screening provided by VFAINT mode was
also applied.  The background light curve extracted from the ACIS-S1
level 2 event file was filtered using the \textsc{lc\_clean}
script\footnote{See http://cxc.harvard.edu/contrib/maxim/acisbg/}
provided by M. Markevitch to identify periods affected by flares.
There were no flares in either observation of Abell 2146 so we
proceeded with the final cleaned exposure of $43\ks$.

As the two separate observations were taken so closely together, with
effectively identical chips positions and roll angles, we were able to
reproject them to a common position (Obs. ID 10464) and combine them.
Exposure-corrected images were created by combining the two cleaned
event files and assuming a monoenergetic distribution of source
photons of $1.5\keV$, which is approximately the peak energy of the
source.

\section{Imaging Analysis}
\label{sec:imganalysis}
\begin{figure*}
\begin{minipage}{\textwidth}
\centering
\includegraphics[width=0.48\columnwidth]{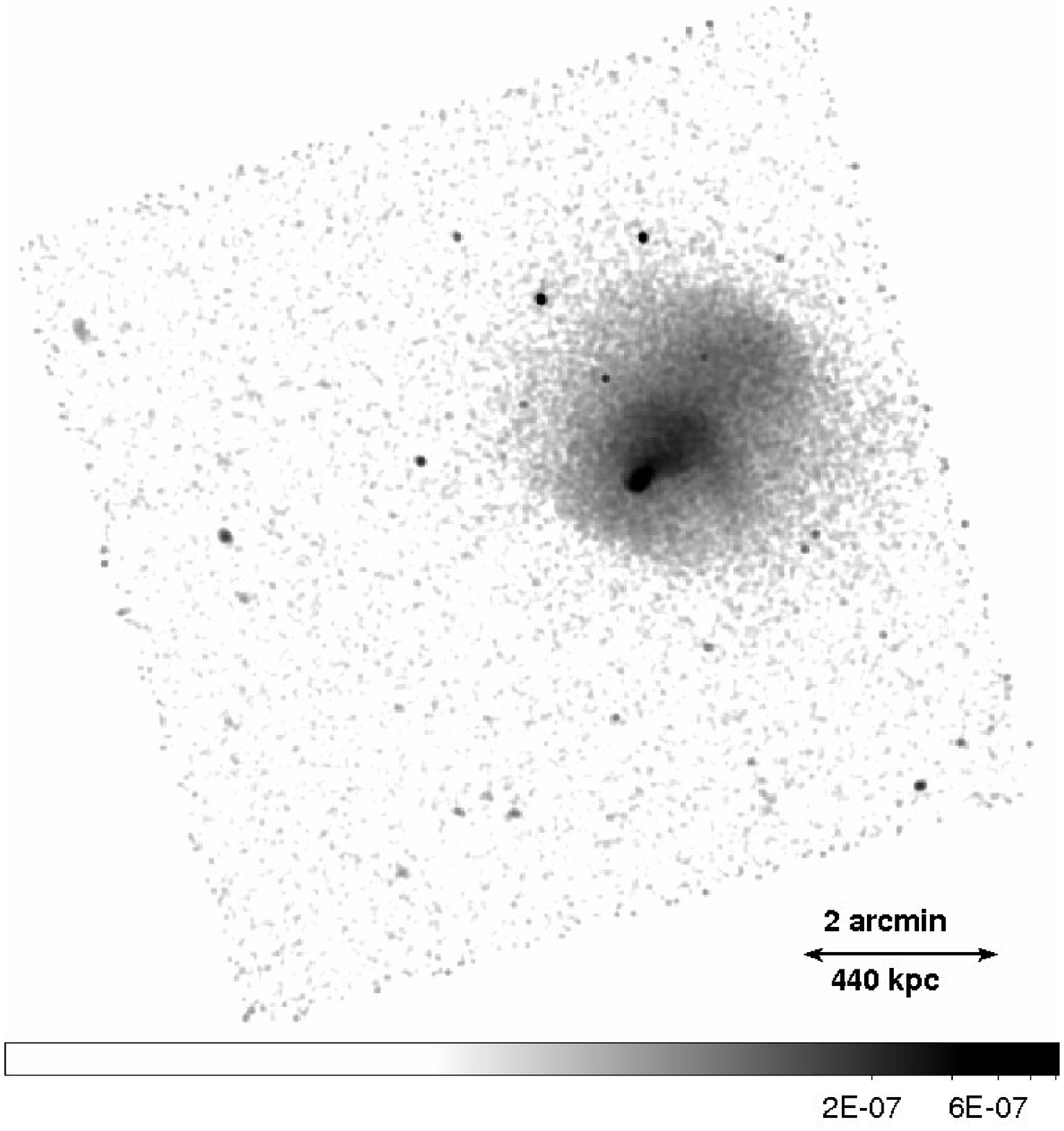}
\includegraphics[width=0.48\columnwidth]{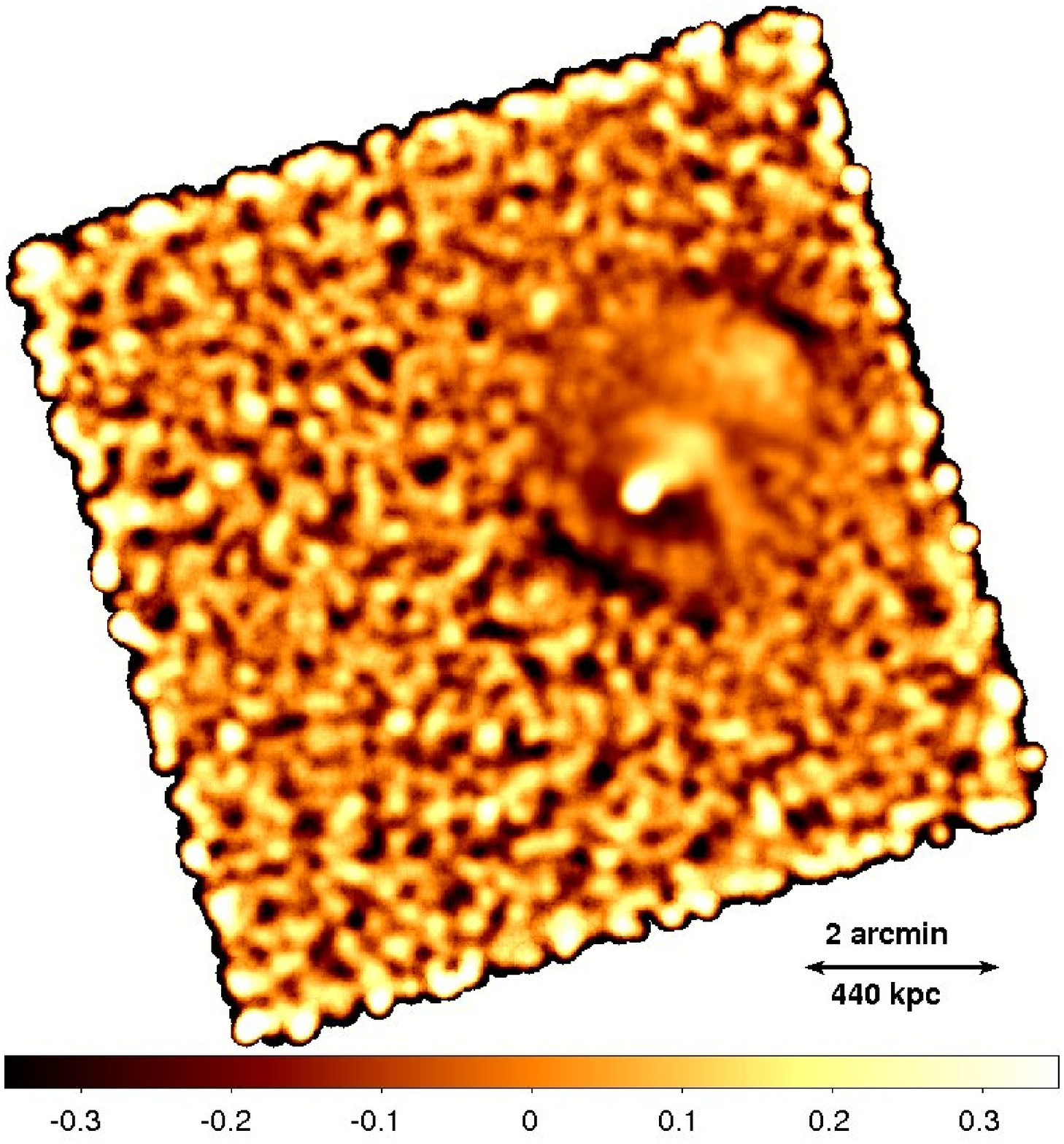}
\caption{Left: Exposure-corrected image in the 0.3--5.0$\keV$ energy
  band smoothed with a 2D Gaussian $\sigma=1.5\asec$ (North is up and
  East is to the left).  The logarithmic scale bar has units
  $\surbri$.  Right: Unsharp-masked image created by subtracting
  images smoothed by 2D Gaussians with $\sigma=5$ and $20\asec$ and
  dividing by the sum of the two images.}
\label{fig:mainimage}
\end{minipage}
\end{figure*}

\begin{figure}
\centering
\includegraphics[width=0.9\columnwidth]{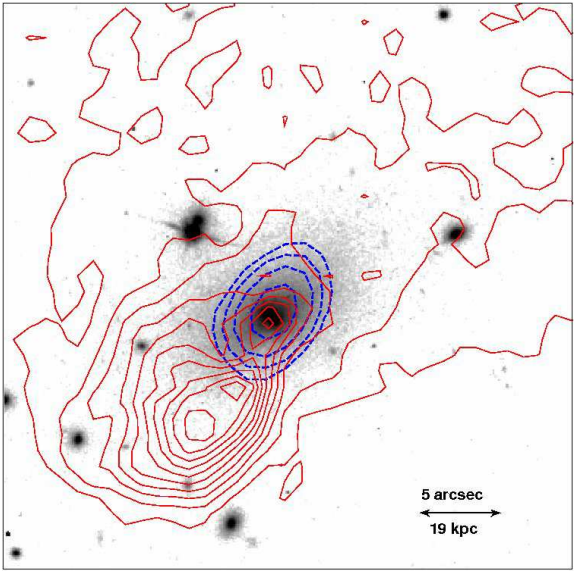}
\caption{Hubble Legacy Archive image of the brightest cluster galaxy
  in Abell 2146 (\citealt{Sand05}) with \emph{Chandra} X-ray and VLA
  $4.9\GHz$ radio (NRAO/VLA Archive Survey) contours superimposed in
  red solid and blue dashed lines respectively.}
\label{fig:zoomimage}
\end{figure}

% Image - main features
Exposure-corrected images of the galaxy cluster and a zoom in of the
cluster core are shown in Fig. \ref{fig:mainimage}.  The X-ray
emission is extended from SE to NW and appears to cut off abruptly at
either end of this axis.  The bright, dense core has been displaced
from the cluster centre and is being stripped of its material to form
a tail of gas towards the NW.  The brightest cluster galaxy (BCG)
shown in Fig. \ref{fig:zoomimage} is displaced $\sim10\asec$ to the NW
of the X-ray surface brightness peak and contains a point source
detected in the \emph{Chandra} X-ray image and a VLA $4.9\GHz$ radio
image (NRAO/VLA Archive Survey).  The X-ray point source is detected
at hard energies ($2-10\keV$) and it is likely that this corresponds
to an AGN, but a measurement of the flux was difficult with the
superimposed cluster emission.  We estimated the point source flux by
extracting the source counts in a region of $2\asec$ radius and
subtracting the cluster emission using a surrounding region from
$2-5\asec$ radius.  Using a powerlaw model with photon index
$\Gamma=2$, we estimated the point source luminosity in the energy
range $2-10\keV$ to be $L_{2-10\keV}=1.6\pm0.4\times10^{42}\ergps$.

\begin{figure}
\centering
\includegraphics[width=0.9\columnwidth]{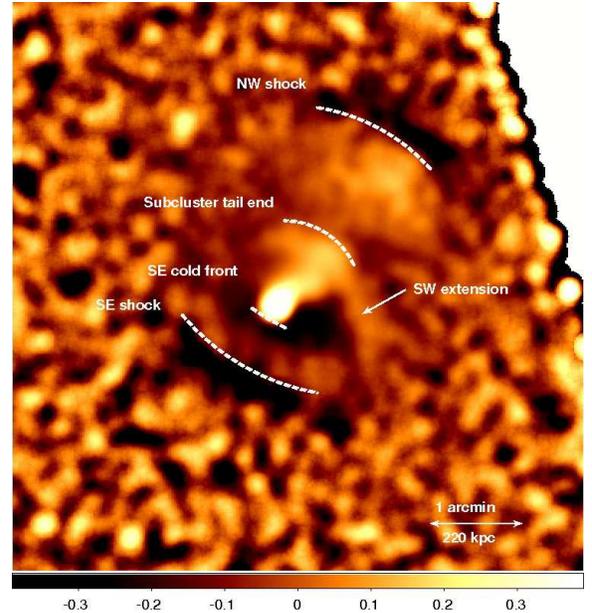}
\caption{Unsharp-masked image as in Fig. \ref{fig:mainimage} with
  structural features in the cluster labelled.}
\label{fig:labelled}
\end{figure}

The unsharp-masked image shown in Fig. \ref{fig:mainimage} (right) highlights
the different structures in the cluster gas.  The cluster is separated into two
concentrations of X-ray emitting gas: the cluster core and its trail
of emission to the SE and a second, more diffuse region of material to
the NW.  Figs. \ref{fig:mainimage} and \ref{fig:labelled} also reveal
several edges in the X-ray surface brightness.  The sharpest of these
defines the SE edge of the bright subcluster core.  A second outer
edge is visible $\sim0.5\arcmin$ to the SE of the core although the
steep decline in the cluster surface brightness reduces the
significance of this feature.  The third edge appears at the NW edge
of the galaxy cluster.  There could potentially be a fourth surface
brightness edge separating the two concentrations of cluster gas.
Finally, the subcluster's tail of gas also appears to have an
extension to the SW.

Abell 2146 appears to have a remarkably similar structure to the
Bullet cluster (\citealt{Markevitch02}; \citealt{Markevitch06}).  The
X-ray morphology suggests a recent merger where a subcluster
containing the dense core has passed through the centre of a second
cluster, the remnant of which appears as the concentration of gas to
the NW.  The dense subcluster has just emerged from the primary core,
travelling to the SE, and is trailing material that has been ram pressure
stripped in the gravitational potential.  The
passage of this dense subcluster core is likely to generate sharp
shock fronts in the ICM.  

\begin{figure}
\centering
\includegraphics[width=0.9\columnwidth]{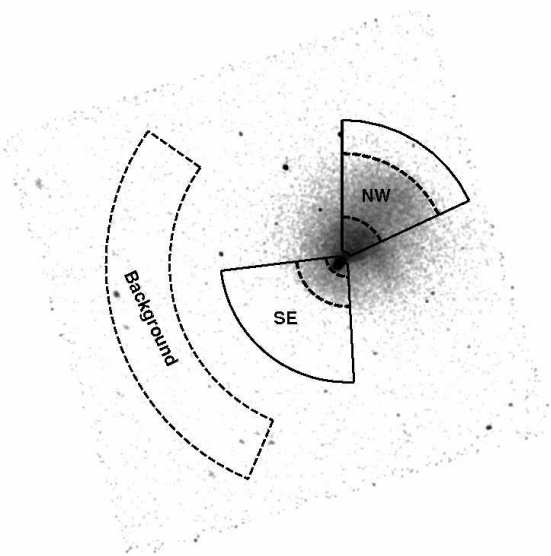}
\caption{Exposure-corrected image of the ACIS-S3 chip
  ($8\times8\amin$) in the 0.3--7.0$\keV$ energy band smoothed with a
  2D Gaussian $\sigma=2.5\asec$.  The sectors and background region
  used to produce surface brightness profiles are labelled.  The
  dashed lines across the sectors mark the approximate location of the
  surface brightness edges.}
% Sectors, shocks, background region
\label{fig:SBsectors}
\end{figure}

To confirm the detection of the edges in the X-ray emission, we
produced surface brightness profiles in two sectors to the NW and SE
from the AGN, as shown in Fig. \ref{fig:SBsectors}.  The surface
brightness profiles were extracted from the merged exposure-corrected
image in the energy range $0.3-7.0\keV$.  The radial bins are $1\asec$
wide in the cluster centre and then increase in size as the cluster
surface brightness declines and the background subtraction becomes
more important.  Point sources were identified using the CIAO
algorithm \textsc{wavdetect}, visually confirmed and excluded from the
analysis using elliptical apertures where the radii were
conservatively set to five times the measured width of the PSF
(\citealt{Freeman02}).  The background was determined in a sector
taken from the SE edge of the ACIS-S3 chip, $200-270\asec$ from the
central AGN, in a region that is largely free of cluster emission
(Fig. \ref{fig:SBsectors}).

% Discuss why circular rather than elliptical apertures were chosen?
% Contour plot?

\begin{figure*}
\begin{minipage}[t]{\textwidth}
\centering
\includegraphics[width=0.48\columnwidth]{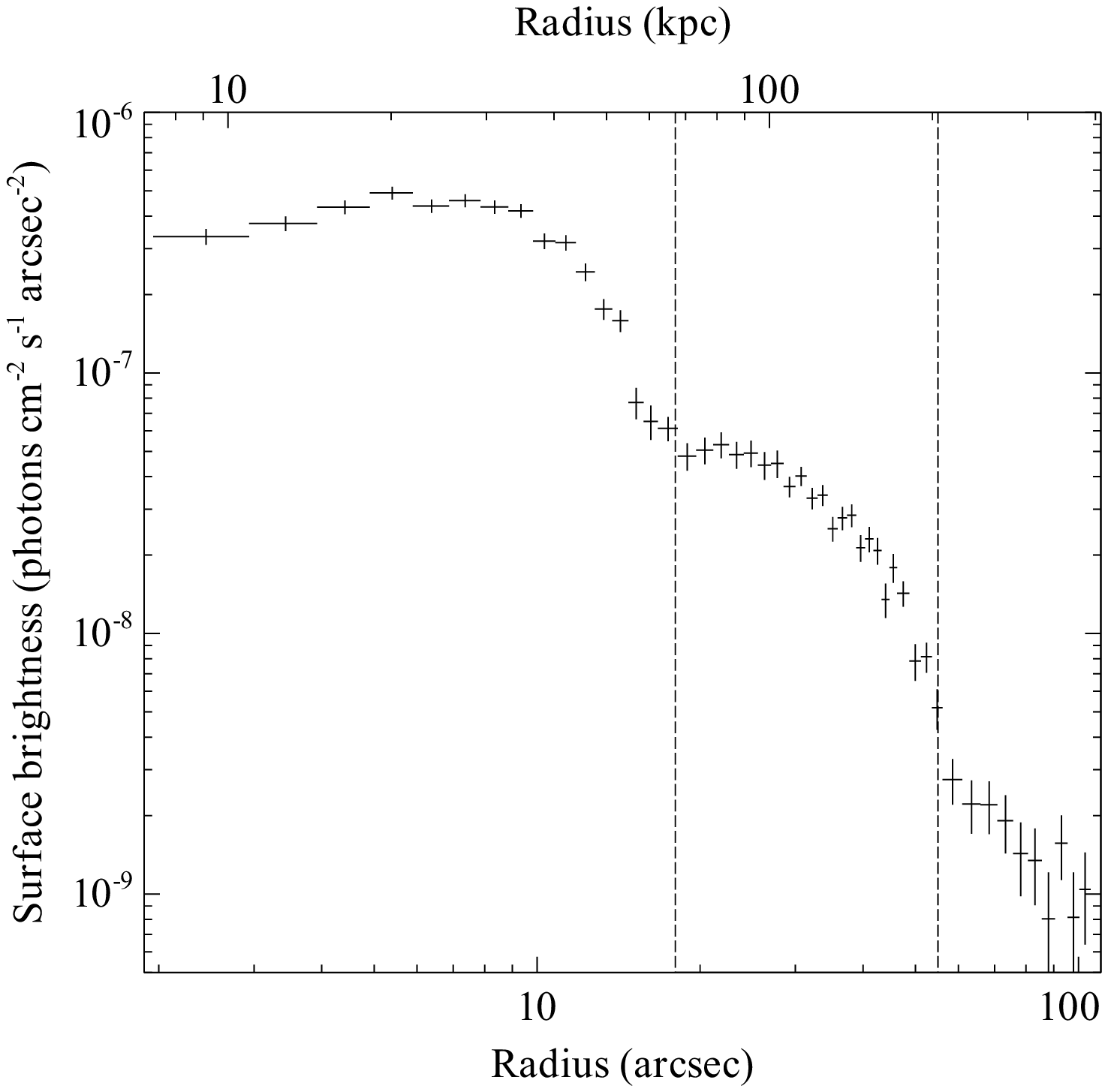}
\includegraphics[width=0.48\columnwidth]{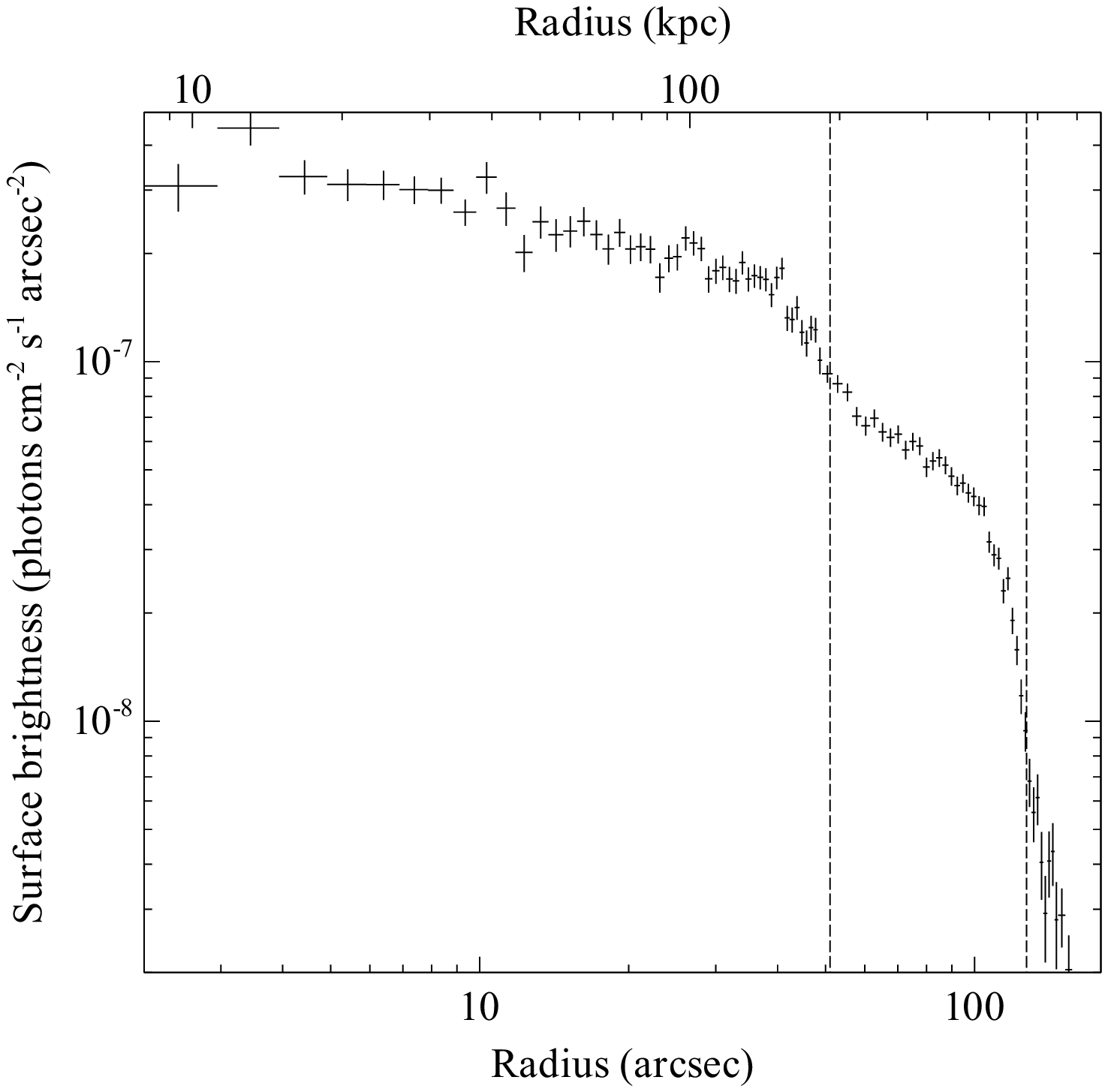}
\caption{Surface brightness profiles in the energy range
  0.3--7.0$\keV$ for the SE (left) and NW (right) sectors as shown in
  Fig. \ref{fig:SBsectors}.  The approximate locations of the surface
  brightness edges shown in Fig. \ref{fig:SBsectors} are marked with
  dashed lines.  Both profiles are centred on the AGN.} % Zero points?
\label{fig:SBprofiles}
\end{minipage}
\end{figure*}

Fig. \ref{fig:SBprofiles} (left) shows the SE surface brightness
profile, centred on the AGN.  There is steep decline in the surface
brightness at $18\asec$ radius, marking the edge of the subcluster
core, and then a second break at $55\asec$.  
%A small increase in the
%surface brightness is also visible at $21\asec$ radius just in front
%of the $18\asec$ edge.  
Fig. \ref{fig:SBprofiles} (right) shows the NW surface brightness
profile declines slowly through the length of the ram pressure
stripped tail to a break at $\sim50\asec$, marking the separation
between the two concentrations of X-ray gas.  There is a sharp edge in
the surface brightness profile at $\sim120\asec$, where the surface
brightness drops by a factor of $\sim7$.  In summary, we identify four
surface brightness edges: two in the SE sector at radii of $18\asec$
and $55\asec$ in front of the subcluster core, and another two in the
NW sector at radii of $50\asec$ and $120\asec$.  To determine whether
these surface brightness edges are shocks or cold fronts
(eg. \citealt{Markevitch00}; \citealt{Vikhlinin01};
\citealt{Markevitch07}), we extracted and analysed X-ray spectra on
either side of these features to determine the ICM temperature and
density.

% Include figure showing SB sectors + marking edges

\section{Spectral Analysis \& Deprojection}
\label{sec:specanalysis}
% Overall spectrum
We first extracted an overall cluster spectrum using an outer radius
of $2\arcmin$, which contained the vast majority of the cluster
emission, after excluding point sources.  The background was
subtracted using a spectrum extracted from a cluster-free region at
the edge of the chip and appropriate responses and
ancillary responses were generated.  The spectrum was restricted to
the energy range $0.5-7.0\keV$ and grouped to contain a minimum of 30
counts per spectral channel.  A single temperature fit to the cluster
spectrum using an absorbed thermal plasma emission model
\textsc{phabs(mekal)} (\citealt{Balucinska92}; \citealt{Mewe85};
\citeyear{Mewe86}; \citealt{Kaastra92}; \citealt{Liedahl95}) in
\textsc{xspec} version 12.5.0 (\citealt{Arnaud96}) produced a good fit
(reduced $\chi^2=0.99$ for 328 degrees of freedom) with a temperature
$6.7^{+0.3}_{-0.2}\keV$, luminosity
$L_{X}=1.55\pm0.02\times10^{45}\ergps$ ($0.01-50.0\keV$) and an
abundance of $0.37\pm0.04\Zsun$, measured assuming the abundance
ratios of \citet{AndersGrevesse89}.  The Galactic hydrogen column
density was left as a free parameter, giving a value
$n_{\mathrm{H}}=0.028\pm0.004\times10^{22}\pcmsq$ which is consistent
with the Galactic value measured by \citet{Kalberla05} of
$n_{\mathrm{H}}=0.03\times10^{22}\pcmsq$.

% Subcluster core
We also fitted an absorbed single temperature model to a spectrum
extracted in a region of radius $10\asec$, which approximately
encloses the bright subcluster core, and excluded the AGN.  The
hydrogen column density parameter was poorly constrained and so was
fixed to the Galactic value.  This produced a temperature of
$2.7\pm0.1\keV$ and a metallicity of $0.6\pm0.1\Zsun$ (reduced
$\chi^2=1.4$ for 68 degrees of freedom).  The spectral fit was
significantly improved by adding a cooling flow \textsc{mkcflow} component which
models gas cooling down to low temperatures ($\chi^2=94$ reduced to
$\chi^2=80$ for 67 degrees of freedom).  The low temperature limit of
the \textsc{mkcflow} model was fixed to $0.1\keV$ and the higher
temperature and metallicity were tied to the \textsc{mekal} component
parameters.  The \textsc{mkcflow} model normalization suggests that
$40\pm10\Msunpyr$ could be cooling out of the X-ray and down to low
temperatures in the subcluster core.

% Discuss fit on L_X-T relation?  Exclude cool-core?
If we exclude the cool core from the overall cluster spectrum we get a
cluster temperature of $7.5\pm0.3\keV$ and luminosity
$L_{X}=1.55\pm0.02\times10^{45}\ergps$ ($0.01-50.0\keV$).  Therefore,
Abell 2146 falls on the $L_x-T$ relation for local clusters
(\citealt{Pratt09}).

\subsection{Temperature and density maps}
\label{sec:contourbin}
\begin{figure*}
\begin{minipage}[t]{\textwidth}
\centering
\includegraphics[width=0.32\columnwidth]{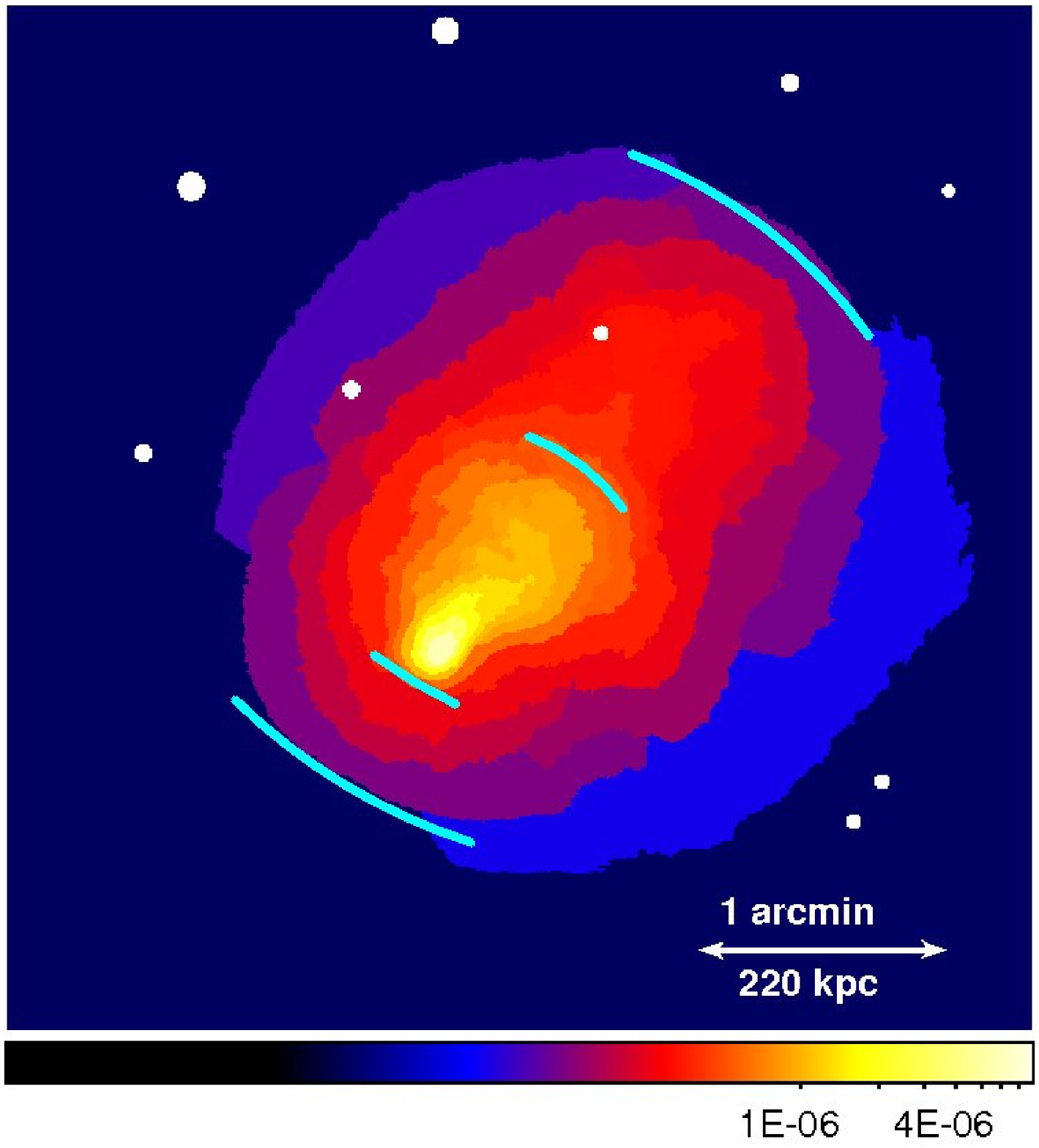}
\includegraphics[width=0.32\columnwidth]{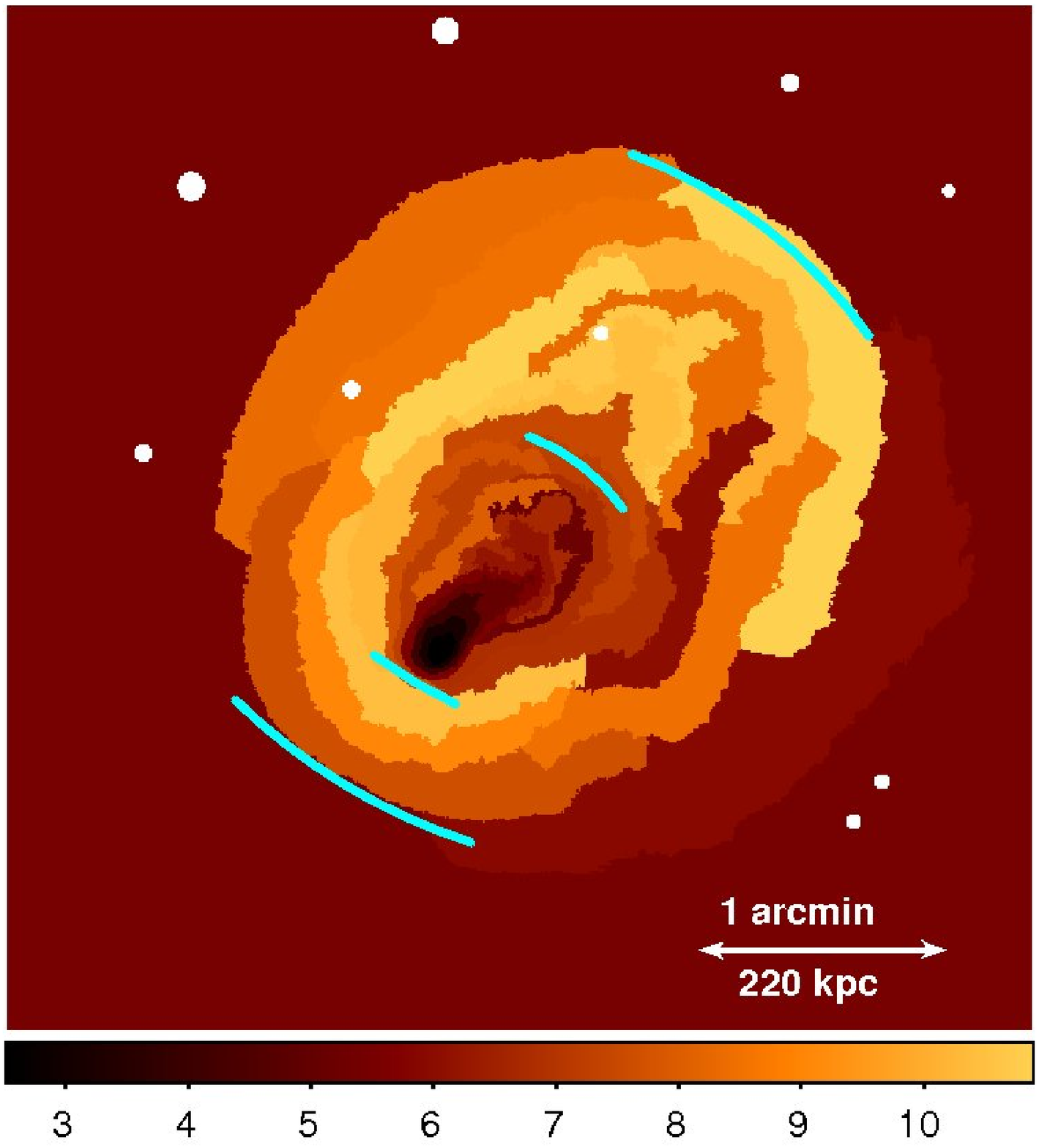}
\includegraphics[width=0.32\columnwidth]{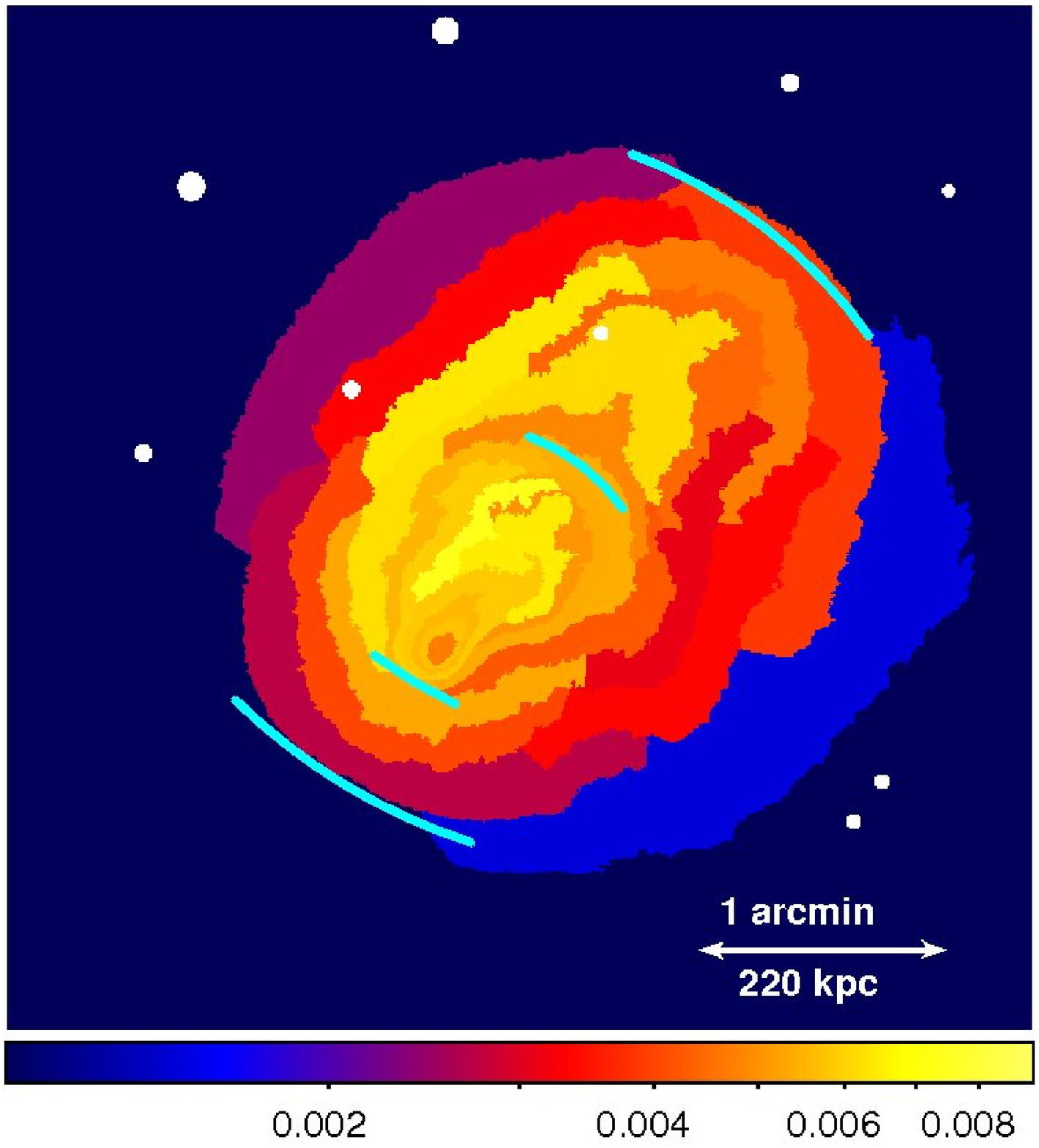}
\caption{Left: projected emission measure per unit area map (units
  $\empasecsq$).  The emission measure is the \textsc{xspec}
  normalization of the \textsc{mekal} spectrum
  $K=EI/(4\times10^{14}\pi D_A^2(1+z)^2)$, where EI is the emission
  integral $EI=\int n_en_H\mathrm{d}V$.  Centre: projected temperature map
  ($\keV$).  Right: projected `pressure' map (units $\keVempasecsq$)
  produced by multiplying the emission measure and temperature maps.
  The blue lines correspond to the dashed lines in
  Fig. \ref{fig:SBprofiles}.  The excluded point sources are visible
  as small white circles.}
\label{fig:jssmaps}
\end{minipage}
\end{figure*}

% Maps of temperature, emission measure
% Discuss background further?
We used spatially resolved spectroscopy techniques to produce maps of
the projected gas properties in the cluster core
(Fig. \ref{fig:jssmaps}).  The central $\sim4\times4\amin$ was
divided into bins using the Contour Binning algorithm
(\citealt{Sanders06}), which follows surface brightness variations.
Regions with a signal-to-noise ratio of 32 ($\sim1000$ counts) were
chosen, with the restriction that the length of the bins was at most
two and a half times their width.  The background spectrum was subtracted from
the observed dataset and appropriate responses and ancillary responses
were generated.  The spectra were grouped to contain a minimum of 20
counts per spectral channel and restricted to the energy range
$0.5-7.0\keV$.  Each spectrum was fitted in \textsc{xspec} with an
absorbed \textsc{mekal} model with the absorption fixed to the
Galactic value $n_{\mathrm{H}}=3.0\times10^{20}\pcmsq$
(\citealt{Kalberla05}) and the redshift fixed to 0.234.  The fitting
procedure minimised the $\chisq$-statistic.  The errors were
approximately $\sim15\%$ in temperature and $\sim8\%$ for the emission
measure.  However, the high temperature bins greater than $10\keV$ are
poorly constrained by the energy range of \emph{Chandra} producing
larger errors greater than $\sim30\%$.  We fixed the metallicity to an
average value of $0.4\Zsun$ in the spectral fits; the limited number
of counts produced a poorly constrained metallicity parameter if it
was left free.  However, we found that fixing the metallicity still produced a
very similar temperature and density map.

% Describe the maps
Shown in Fig. \ref{fig:jssmaps} are the projected emission measure per unit
area, temperature and `pressure' maps.  The projected `pressure' map was
produced by multiplying the square root of the emission measure per
unit area and the temperature maps.  The projected emission measure map shows
the strongly peaked core surface brightness, bright tail of stripped
material and the elongated morphology of the cluster in the NW to SE
direction.  There could also be a spur of emission out to the SW of
the subcluster core.  The projected temperature map shows strong variations
across the cluster.  In the dense subcluster core, the temperature
drops as low as $1.9\pm0.1\keV$ and then sharply increases to the SE
up to $8-10\keV$ between the two SE surface brightness edges.  The
large errors on the high temperature $10^{+3}_{-2}\keV$ bin
immediately in front of the cool subcluster make it consistent with a
constant temperature between the edges.  The ram pressure stripped
tail of material appears as a warmer stream of gas ($5-8\keV$) behind
the subcluster core and trails back to the hottest region of the
disrupted main cluster.  The SW spur of emission suggested in the
emission measure map corresponds to a $5-7\keV$ region of gas which
could be connected to the ram pressure stripped tail from the core.
The high temperatures, $11^{+3}_{-2}\keV$, in the NW region of the
cluster suggest shock heating of the gas is likely.  The sudden drop
in temperature at the NW edge of the cluster, which coincides with the
surface brightness edge, is particularly suggestive of a shock front.

The NW edge is clearly visible in the projected `pressure' map
(Fig. \ref{fig:jssmaps} right) as a sudden increase and
subsequent drop in the pressure by a factor of $\sim5$.  To the SE of
the subcluster core, the pressure is approximately constant for
$\sim20\asec$ and then drops abruptly at larger radii.  This could
indicate that a contact discontinuity or cold front immediately
precedes the subcluster rather than a shock (\citealt{Markevitch00};
\citealt{Vikhlinin01}; \citealt{Markevitch07}).

% Discuss the metallicity map

\subsection{Projected \& deprojected radial profiles}
\begin{figure}
\centering
\includegraphics[width=0.98\columnwidth]{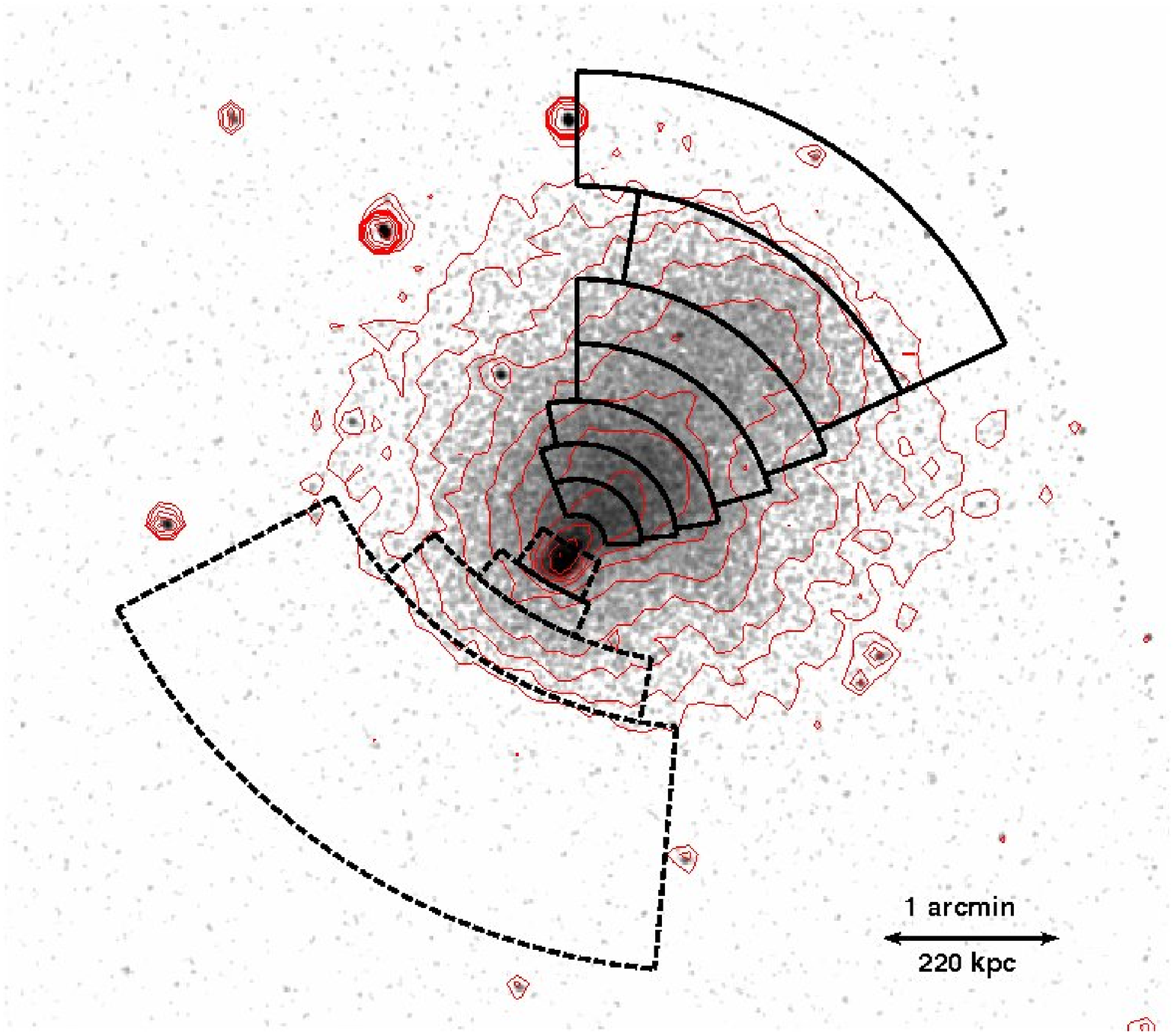}
\caption{Exposure-corrected image in the 0.3--7.0$\keV$ energy band
  smoothed with a 2D Gaussian $\sigma=2.5\asec$.  X-ray surface
  brightness contours have been overlaid (thin red solid lines).  The regions used to analyse the
  NW and SE surface brightness edges are shown by thick black solid
  and dashed lines respectively.}
\label{fig:deprojsectors}
\end{figure}

% More significant detection using spectra in sectors
We obtained a more significant detection of the temperature and
density changes across the surface brightness edges by extracting
radial profiles in the NW and SE sectors
(Fig. \ref{fig:deprojsectors}).  The radial bins were positioned so as
to determine the gas properties on either side of the surface
brightness edges while maintaining a minimum of 3000 counts in each
extracted spectrum.  This lower limit ensured enough counts to provide
a good spectral fit and constraints on the parameters.  The steep
decline in the cluster surface brightness in the outermost radial bins
made it particularly difficult to constrain the outer gas properties.
A wide outer radial bin from $1$ to $2.5\amin$ was therefore required
for the SE sector to determine the nature of the surface brightness
edge in front of the dense core.

% Profiles in different sectors

%% Note in here about blank sky backgrounds
Spectra were extracted from the radial bins in each sector.  Point
sources were identified and excluded as detailed in section
\ref{sec:imganalysis}.  The spectra were analysed in the energy range
$0.5-7.0\keV$ and grouped with a minimum of 30 counts per spectral
bin.  The background subtraction was particularly important for the
regions outside the surface brightness edges where the flux is low.
We compared the use of an on-chip background extracted from a
source-free sector at the edge of the chip (Fig. \ref{fig:SBsectors})
with a blank-sky background extracted from the data sets available
from the CXC and normalized to the count rate in the high energy band
$9.5-12\keV$.  Although the use of a blank-sky background would
account for spatial variations in the background count rate across the
chip, this was estimated to be only a few per cent in the energy band
$0.5-7.0\keV$.  The differences between the blank-sky and on-chip
backgrounds when extracted from the same source-free region were found
to be more significant.  The normalized blank-sky spectrum overestimated the
background at low energies $<1\keV$ and underestimated the background
count rate in the energy range $2-5\keV$ by a factor of $\sim1.3$
which significantly affected the temperature result in regions of low
surface brightness.  We therefore used a background spectrum extracted
from the cluster-free sector at the edge of the chip for the spectral
analysis.  Response and ancillary response files were generated for
each spectrum, weighted according to the number of counts between 0.5
and $7.0\keV$.  These projected spectra were then fitted in
\textsc{xspec} with an absorbed single temperature thermal plasma
emission model \textsc{phabs(mekal)}.  The redshift was fixed to
$z=0.234$ and the absorption was fixed to the Galactic value
$n_{\mathrm{H}}=0.03\times10^{22}\pcmsq$ (\citealt{Kalberla05}).
% Discuss n_H value + abundance here

% Discuss deprojection for such an irregular object => compare
% projected + deprojected results
% Discuss assumption of spherical symmetry
However, these projected spectra extracted from the cluster on the plane of the
sky correspond to summed cross-sections of all the emission along the
line of sight.  A spectrum from the centre of a relaxed cluster will
therefore contain a range of spectral components from the core to the
cluster outskirts.  To determine the properties of the cluster core
these projected contributions from the outer cluster layers should be
subtracted off the inner spectra by deprojecting the emission.
However, deprojection routines require information about the line of
sight extent of the cluster and generally assume that the cluster is
spherically symmetric.  While this is a reasonable assumption for a
relaxed cluster, the highly irregular and elongated morphology of
Abell 2146 clearly deviates from spherical symmetry.  We have
therefore compared the projected and deprojected spectra and discussed
the validity of the assumption of spherical symmetry for each sector
considered.  % more detail here?

We used a straightforward model-independent spectral deprojection
routine (\textsc{dsdeproj}; \citealt{SandersFabian07};
\citealt{Russell08}), which assumes spherical symmetry.
\textsc{dsdeproj} starts from the background-subtracted spectrum
extracted from the outermost annulus and assumes it was emitted from
part of a spherical shell.  This spectrum is scaled by the volume that
is projected onto the next innermost shell (geometric factors from
\citealt{Kriss83}) and subtracted from the spectrum extracted from
that annulus.  In this way the deprojection routine moves inwards
subtracting the contribution of projected spectra from each successive
annulus to produce a set of deprojected spectra.

The deprojected spectra were also analysed in the energy range
$0.5-7.0\keV$ and grouped with a minimum of 30 counts per spectral
bin.  Appropriate response and ancillary response files were
generated as before.  The deprojected spectra were fitted in
\textsc{xspec} with an absorbed \textsc{mekal} model and the
parameters set as previously described for the projected spectra.

\subsection{NW sector: upstream shock}
\label{sec:NWsector}

\begin{figure}
\centering
\includegraphics[width=0.98\columnwidth]{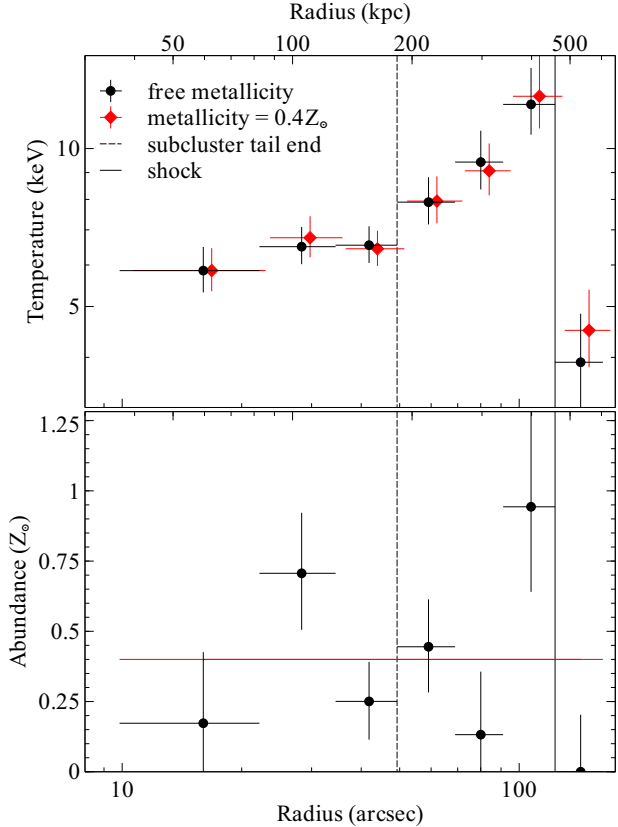}
\caption{NW sector projected radial temperature (upper) and
  metallicity (lower) profiles centred on the AGN.  The black circles
  show the radial temperature profiles with the metallicity left as a
  free parameter and the red diamonds show the results when the
  metallicity was fixed to $0.4\Zsun$ (slightly offset in the
  x-direction for clarity).}
\label{fig:NWsectorproj}
\end{figure}

\begin{figure}
\centering
\includegraphics[width=0.98\columnwidth]{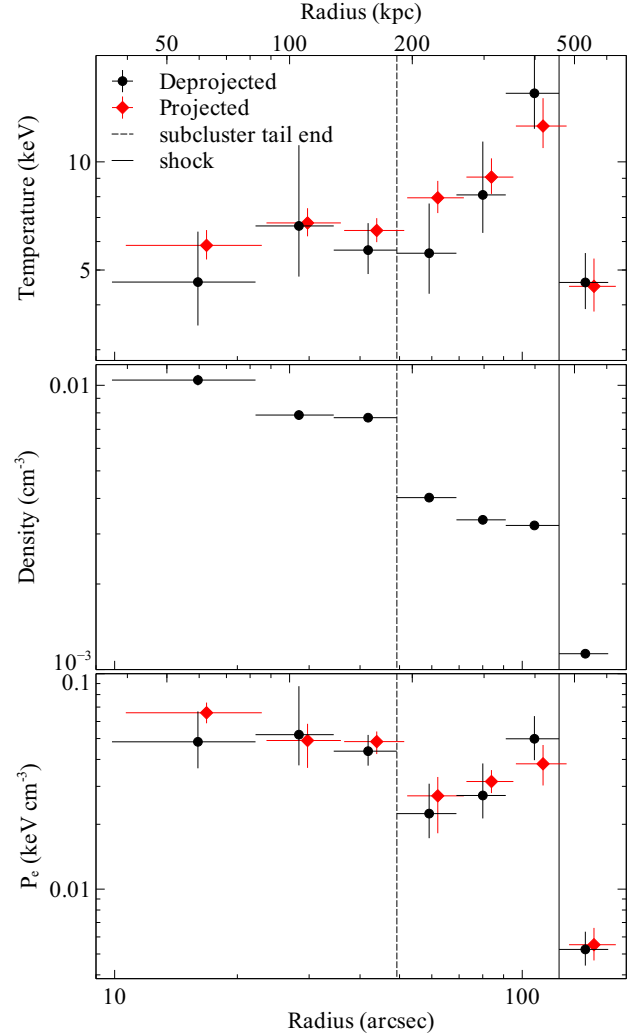}
\caption{NW sector deprojected radial temperature (upper), electron
  density (centre) and electron pressure (lower) profiles centred on
  the AGN.  The metallicity was fixed to $0.4\Zsun$.  The projected
  temperature profile, with metallicity also fixed to $0.4\Zsun$, is
  shown overlaid on the deprojected profile (slightly offset in the
  x-direction for clarity).  The projected electron
  pressure was calculated using the projected temperatures and
  deprojected electron densities.  The dashed line marks the tail end
  of the subcluster and the solid line marks the upstream shock.}
\label{fig:NWsectordeproj}
\end{figure}

Figs. \ref{fig:NWsectorproj} and \ref{fig:NWsectordeproj} show the
projected and deprojected radial profiles for the NW sector of Abell
2146.  The projected temperature is approximately constant at
$\sim6.5\keV$ through the length of the ram pressure stripped trail of gas
from the cool-core.  Beyond a radius of $50\asec$ ($200\kpc$) from the central
AGN, marking the approximate end of the subcluster tail, the temperature
increases steadily to a peak of $13^{+2}_{-2}\keV$ at $100\asec$ ($400\kpc$).  Then
at the surface brightness edge the temperature plummets down to
$4.5^{+0.9}_{-0.7}\keV$.  The metallicity parameter was not well
constrained by the spectral fits and the increase seen at $100\asec$ is
not significant.  The sharp drop in the temperature at this radius is
matched by a drop in the electron density shown by the deprojected
profiles in Fig. \ref{fig:NWsectordeproj} confirming that this is a
shock.

% relevance of deprojection - outer bin unlikely to have projected
% emission => at background level here.
The deprojection routine \textsc{dsdeproj} assumes spherical symmetry,
which may not be reasonable given the cluster's highly elongated
morphology in this sector.  However, the steep gradient of the surface
brightness profile outside $40\asec$ radius
(Fig. \ref{fig:SBprofiles}) reduced the significance of the projected
outer layers.  The deprojected temperatures were therefore consistent
with the projected temperatures to within the $1\sigma$ errors (Figs.
\ref{fig:jssmaps} and \ref{fig:NWsectorproj}).  We used the
deprojected results to analyse the shock at $120\asec$ where the
surface brightness contours tend toward circular at the shock edge.  The
elongated morphology and substructure of the subcluster tail, coupled
with the shallow surface brightness gradient, made the deprojection of
the surface brightness edge at $50\asec$ much more difficult.  For a
good constraint on the temperature, a much greater number of X-ray
counts is needed to facilitate the use of smaller radial bins in a narrower NW
sector.  The metallicity was poorly constrained
in the deprojected annuli therefore this parameter was fixed to the
average of $0.4\Zsun$ determined from the spectral fits to the
projected annuli.

% Discuss deprojected result
There are two sharp drops in the deprojected electron density profile
(Fig. \ref{fig:NWsectordeproj}) which correspond to the outer surface
brightness edge at $120\asec$ seen in Fig. \ref{fig:SBprofiles}
(right) and the subcluster tail end at $50\asec$.  At a radius of
$120\asec$, the density decreases by a factor of $2.83\pm0.08$,
which coincides with a temperature drop from $16^{+4}_{-3}\keV$ down
to $4.6^{+1.0}_{-0.7}\keV$ confirming that this outer edge is a shock.  By
combining the deprojected temperature $T$ and electron density $n_e$,
we calculated the electron pressure $P_e=k_Bn_eT$
(Fig. \ref{fig:NWsectordeproj} bottom).  As expected, there is a
large decrease in pressure by factor of $10^{+3}_{-2}$ at the shock
front.

% Need to also discuss subcluster tail end!
The nature of the edge at the subcluster tail end ($50\asec$) was more
difficult to determine.  The radial temperature bin outside the edge
($50-70\asec$) is likely to contain some hotter gas from the outer
shock edge in projection, biasing the projected temperature value
high.  The deprojected temperature in this radial bin is poorly
constrained so it is unclear whether the subcluster tail ends in a
shock or a cold front.  The density decreases by a factor of
$1.91\pm0.07$ producing a drop in the electron pressure by a factor of
$2.1^{+0.8}_{-0.9}$.  The decrease in pressure across the surface
brightness edge was not very significant primarily because of the poor
constraints on the gas temperature.  

% Calculate Mach no.
Following \citet{LandauLifshitz59}, we applied the Rankine-Hugoniot
jump conditions across the shock at $120\asec$ radius to calculate the
Mach number, $M=v/c_s$, where $v$ is the velocity of the pre-shock gas
and $c_s$ is the velocity of sound in that gas.  The Mach number can
be calculated independently from the density jump,

\begin{equation}
M=\left(\frac{2 \ ^{\rho_2}/_{\rho_1}}{\gamma + 1 - \
    ^{\rho_2}/_{\rho_1}(\gamma -1)}\right)^{1/2}
\label{eq:Mdensity}
\end{equation}

\noindent or temperature jump,

\begin{equation}
M=\left(\frac{(\gamma + 1)^2 \left(^{T_2}/_{T_1} - 1 \right)}{2\gamma(\gamma-1)}\right)^{1/2}
\label{eq:Mtemp}
\end{equation}

\noindent where $T_1$, $\rho_1$ and $T_2$, $\rho_2$ denote the
temperature and density before/upstream and after/downstream of the
shock respectively.  Here we assume the adiabatic index for a
monatomic gas, $\gamma=5/3$.  This may not be applicable if, for
example, a significant amount of energy is lost in the acceleration of
particles at the shock front.

% Second equation assumes temperature equilibrium between ions +
% electrons (because shock affects ion temperature + we measure
% electron temp.)

At the NW shock, the density drops by a factor
$\rho_2/\rho_1=2.83\pm0.08$ which, from equation \ref{eq:Mdensity},
gives a Mach number $M=2.7\pm0.1$.  Using equation \ref{eq:Mtemp}, the
observed temperature drop of $T_2/T_1=3.4^{+1.2}_{-0.9}$ gives
$M=2.7^{+0.7}_{-0.5}$.  These two independent calculations of the Mach
number agree within the $1\sigma$ errors.

However, the sharp drop in surface brightness produced by the shock
edge is superimposed on the underlying decay in the cluster surface
brightness with radius.  The relatively large radial bins used in this
analysis to ensure a reliable calculation of temperature may therefore
overestimate the density drop at the shock.  If free-free emission
dominates, the X-ray emissivity depends strongly on gas density and
only weakly on temperature, $\varepsilon \propto \rho^2T^{1/2} $
permitting narrower radial bins for calculating the density.  Since
the ICM temperature in Abell 2146 only drops below $2\keV$ in the cool
subcluster core this is a reasonable assertion.  We therefore
deprojected the X-ray surface brightness profile, using small
corrections for the temperature variation, and derived the electron
density in narrower radial bins to more accurately calculate the
density jumps across the shock edges.

The surface brightness profile was deprojected by assuming spherical
symmetry and using the straightforward `onion-peeling' method first
described in \citet{Fabian80}.  The background was subtracted using
the region at large radii from the cluster as before
(Fig. \ref{fig:SBsectors}).  The radial bins are slightly larger for
the deprojected profile to ensure a similar number of counts at each
radius.  However, there were few data points above the background
level outside the NW shock edge to constrain the upstream gas
properties.  Although there are \emph{ROSAT} observations of Abell
2146, in the PSPC exposures of the field it is more than $40\amin$ off
axis and the HRI exposure of the cluster on axis does not detect the
cluster outskirts with any significance.  It was therefore not
possible to verify the analysis of the outer cluster gas layers with
\emph{ROSAT} data.

The deprojected electron density was calculated from the
\textsc{xspec} normalization for a \textsc{phabs(mekal)} model, where
the total model flux was set equal to the deprojected surface
brightness in each radial bin.  The temperature parameter was set to
the corresponding deprojected value determined from spectral fitting
(Fig. \ref{fig:NWsectordeproj} upper).

\begin{figure}
\centering
\includegraphics[width=\columnwidth]{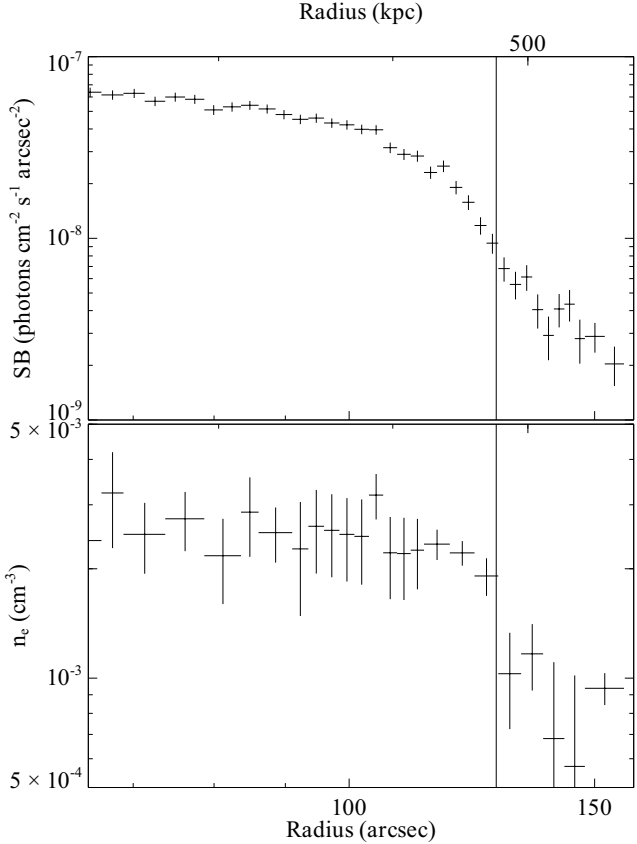}
\caption{Upper: Surface brightness profile in the energy range
  0.3--7.0$\keV$ for the NW shock.  Lower: deprojected electron
  density profile.  The shock edge is marked with a solid line.}
\label{fig:NWshockSB}
\end{figure}

The projected surface brightness and deprojected electron density
across the NW shock edge are shown in Fig. \ref{fig:NWshockSB}.  
%Note that the increase in the outermost radial bin of the electron
%density profile is an artifact of the deprojection process.The X-ray
%emission extends beyond the outermost annulus so that the emitting
%volume associated with this annulus is too small.  This causes an
%overestimate of the projection on to the next annulus in so that the
%emission in that radial bin is underestimated.  However, the effect is
%minimal for annuli further in as the steep surface brightness limits
%the amount of projection.  
Fig. \ref{fig:NWshockSB} (lower panel)
shows a drop in the gas density of $\rho_2/\rho_1=2.0\pm0.4$ across
the shock edge.  From equation \ref{eq:Mdensity}, this gives a Mach
number $M=1.7\pm0.3$, lower than the spectral fitting result
calculated from the density jump of $M=2.7\pm0.1$.  If we repeat the spectral
deprojection, but with each radial bin split into two and
the temperature parameters tied together in each pair, the Mach number
calculated from the density also drops to $M=2.1^{+0.2}_{-0.1}$.

We therefore conclude that the Mach number of the NW shock is closer
to the lower value of $M=1.7\pm0.3$ calculated from the higher
resolution spatial binning which more closely probes the shock.  The
sound speed in front of the shock is $c_s=(\gamma
k_BT_1/m_H\mu)^{1/2}=1100^{+100}_{-90}\kmps$, where the gas temperature
in front of the shock $T_1=4.6^{+1.0}_{-0.7}\keV$ and the mean molecular
weight of the medium $\mu=0.6$.  Therefore, for a Mach number of
$M=1.7\pm0.3$ the shock velocity is $v=Mc_s=1900\pm400\kmps$.

\subsection{SE sector: cold front \& bow shock}
\label{sec:SEsector}
\begin{figure}
\centering
\includegraphics[width=0.98\columnwidth]{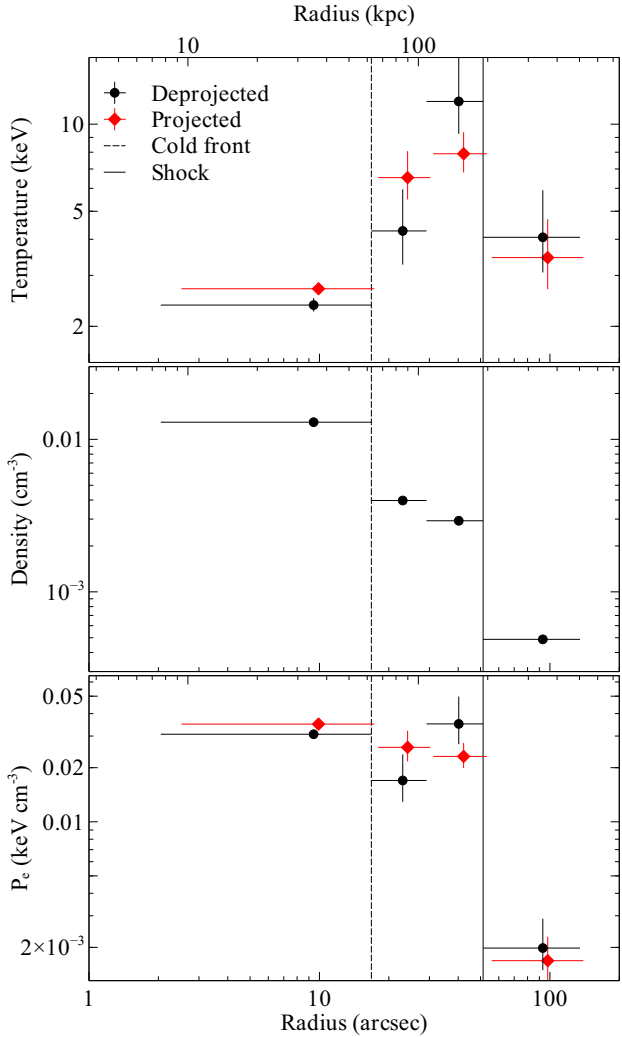}
\caption{SE sector projected and deprojected radial temperature
  (upper), electron density (centre) and electron pressure (lower)
  profiles centred on the AGN.  The metallicity was fixed to
  $0.4Z_{\odot}$.  The projected electron pressure was calculated
  using the projected temperatures and deprojected electron densities.
  The projected data points are slightly offset in the x-direction for
  clarity.}
\label{fig:SEsector}
\end{figure}

The projected and deprojected radial profiles in temperature, density
and electron pressure for the SE sector in front of the subcluster
core are shown in Fig. \ref{fig:SEsector}.  The assumption of
spherical symmetry applied by \textsc{dsdeproj} was considered to be
reasonable as the cluster appears approximately circular on the sky
within this sector and the radial bins traced the surface brightness
contours (Fig. \ref{fig:deprojsectors}).  The metallicity was poorly
constrained in the spectral fits and so was fixed to an average value
of $0.4\Zsun$ in this sector.

The subcluster core, analysed by the innermost radial bin, contains
the lowest temperature, $2.4\pm0.1\keV$, and highest density gas,
$n_e=0.0130\pm0.0001\pcmcu$, in the cluster.  The radiative cooling
time of the gas $t_{\mathrm{cool}}$ was derived from the temperature $T$ and density $n_e$

\begin{equation}
t_{\mathrm{cool}} = \frac{5}{2}\frac{nk_{B}T}{n_en_H\Lambda(T)}
\end{equation}

\noindent where $\Lambda(T)$ is the cooling function, $n$ is the total number
density of gas particles and $n_H$ is the number density of hydrogen.
In the subcluster core the radiative gas cooling time drops to
$3.5\pm0.2\Gyr$.  Therefore it is likely to be a cool core remnant
that is being ram pressure stripped in the merger.

The temperature of the gas increases rapidly in front of the
subcluster core by a factor of $1.8^{+0.7}_{-0.4}$.  This jump
coincides with the surface brightness edge at $18\asec$ visible in
Fig. \ref{fig:SBprofiles} and a drop in the electron density by a
factor of $3.3^{+0.1}_{-0.1}$.  The electron pressure is therefore
approximately constant across this surface brightness edge.  This edge
is a cold front or contact discontinuity, similar to that found at the
leading edge of the bullet in the Bullet cluster
(\citealt{Markevitch02}) and in other clusters such as Abell 2142
(\citealt{Markevitch00}) and Abell 3667 (\citealt{Vikhlinin01}).

The gas properties at the outer edge at $55\asec$ radius were more
difficult to extract.  The sharp drop in the cluster surface
brightness necessitated a large outermost radial bin, which therefore
did not allow constraints on the properties close to the edge.  There
is a very sharp drop in the electron pressure by a factor of
$18^{+11}_{-6}$ at $55\asec$ corresponding to a decrease by factors of
$3^{+2}_{-1}$ and $6.0^{+0.3}_{-0.2}$ in temperature and density
respectively.  However, the Rankine-Hugoniot jump conditions
(eq. \ref{eq:Mdensity}) give a maximum possible density contrast
across adiabatic shocks in a monatomic gas of a factor of four.  This
limit arises because as the strength of the shock is increased to
higher Mach numbers the high thermal pressure building behind the
shock limits the compression of the post-shock gas.  The large radial
bins ($40\pm11\asec$ and $93\pm42\asec$) used for this analysis do not
closely probe the gas properties across the shock edge.  The
underlying decline in the cluster density profile, which is
significant in such a large radial region, is included in the
calculation of the density change at the shock.  This produces an
overestimate of the density jump.

\begin{figure}
\centering
\includegraphics[width=\columnwidth]{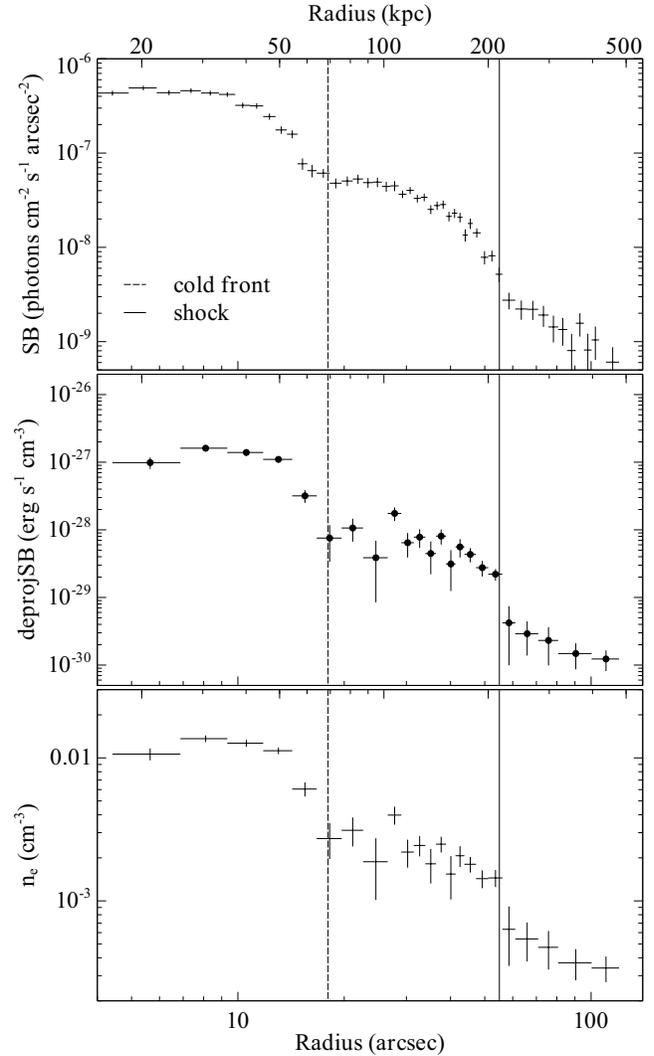}
\caption{Upper: Surface brightness profile in the energy range
  0.3--7.0$\keV$ for the SE shock.  Centre:
  deprojected surface brightness profile Lower: deprojected electron
  density profile with bestfit values on either side of the shock
  edge.  The dotted line marks a possible density enhancement in the
  stagnation region in front of the subcluster.}
\label{fig:SEshockSB}
\end{figure}

We have therefore also calculated the density jump by deprojecting the
surface brightness profile across the shock.  Fig. \ref{fig:SEshockSB}
shows the drop in surface brightness across the shock edge at
$55\asec$ radius corresponds to a drop in the gas density by a much
lower factor of $\rho_2/\rho_1=2.4\pm0.7$.  Using equation
\ref{eq:Mdensity}, the Mach number of the SE bow shock is therefore
$M=2.2\pm0.8$.  For a preshock temperature of $4^{+2}_{-1}\keV$, which
gives a sound speed of $c_s=1000^{+200}_{-100}\kmps$, the shock
velocity is $v=2200^{+1000}_{-900}\kmps$.  The subcluster is at a
projected distance of roughly $380\kpc$ from the centre of the main
cluster and, if we assume that the subcluster moves at the shock
velocity (but see \citealt{Springel07}), this implies that the
subcluster passed through the centre of the main cluster approximately
$0.1-0.3\Gyr$ ago.  

\section{Discussion}
% Formation scenarios discussion, relate Mach no., upstream shock +
By analysing the temperature and density of the gas on either side of
the surface brightness edges, we have determined that the merging
cluster Abell 2146 contains two shock fronts: a bow shock in front of
the subcluster with $M=2.2\pm0.8$ and a slower upstream shock behind
the main cluster with $M=1.7\pm0.3$.  We estimate that the subcluster
passed through the centre of the main cluster approximately
$0.1-0.3\Gyr$ ago and is being ram pressure stripped of its material.
The sharpness of the observed surface brightness edges
(Fig. \ref{fig:SBprofiles}) suggests that Abell 2146 is oriented close
to the plane of the sky.  However, a measurement of the line of sight
velocities for the subcluster and the main cluster galaxies using
optical spectroscopy, combined with the Mach numbers, will give a
quantitative constraint on the inclination of the merger axis.  

% Luminosity + temperature boost
Although Abell 2146 is a smaller system with a much lower global
temperature, it appears remarkably similar in structure to the Bullet
cluster (\citealt{Markevitch02}; \citealt{Markevitch06}).  Shock
fronts are so rarely detected because the merger must be observed
before the shock has moved to the outer, low surface brightness
regions of the cluster and with a merger axis close to the plane of
the sky so that projection effects do not conceal the surface
brightness edges.  It is therefore unsurprising that Abell 2146 appears
similar to the Bullet cluster and is observed at a comparable time since core
passage of the subcluster (both around $\sim0.2\Gyr$).  

By choosing the brightest clusters to observe we will also
preferentially select merging clusters in this short window of core
passage.  Merging clusters undergo a dramatic increase in luminosity
and temperature as the cores collide, pushing them up the $L_x-T$
relation (\citealt{Ricker01}; \citealt{Ritchie02};
\citealt{Markevitch06}).  This boost is caused by the compression of
the cluster cores during the merger and can be up to a factor of 10
for the luminosity in a head-on collision between two equal mass
clusters.  Statistically more likely collisions between clusters with
higher mass ratios of 3:1 and 8:1 and low impact parameters (less than
a few core radii) will still produce an increase in luminosity by a
factor of $\sim4$ and $\sim2$, respectively (\citealt{Ricker01};
\citealt{Ritchie02}; \citealt{Poole07}).  However, this strong
variation is relatively short-lived and disappears after approximately
one sound crossing time.  For a typical cool-core cluster, with a core
radius of $100\kpc$ and temperature $3\keV$, the sound crossing time
is only $0.1\Gyr$.

% Mention shocks are more easily observed here?

\subsection{Mass ratio of the merging clusters}
% Upstream shock - discuss simulations
The key morphological difference between Abell 2146 and the Bullet
cluster is the clear detection of an upstream shock.  An upstream shock is
generated as the gravitational potential minimum fluctuates rapidly
during core passage (see eg. \citealt{Roettiger97}; \citealt{Gomez02}).  The
gravitational potential reaches an extreme minimum as the two cluster
cores coalesce.  This causes a significant amount of the outer cluster
gas to flow inwards.  The subcluster then exits the main core and the
gravitational potential rapidly returns to its premerger level, which
expels much of the newly arrived gas.  The gas that is expelled in the
upstream direction then collides with the residual infall from the
subcluster and forms a shock propagating in the opposite direction to
the subcluster.

% Mass ratio arguments
For a higher mass ratio merger, such as the Bullet cluster (10:1;
\citealt{Clowe04}; \citealt{Bradac06}; \citealt{Clowe06}), the
perturbation to the gravitational potential caused by the merger is
smaller relative to the total potential and shorter in duration.  The
outer gas layers have less time to respond to the change in
gravitational potential, which reduces the infall and strength of the
subsequent outflow, producing only a weak upstream shock.  In addition,
there is relatively little residual infall from the wake of the small
subcluster.  Simulations of cluster mergers over a range of mass
ratios have shown that the more dramatic variation in the
gravitational potential during lower mass ratio mergers (less than
8:1; \citealt{Roettiger97}) produces a stronger upstream shock
(\citealt{Roettiger97}; \citealt{Ricker01}; \citealt{Poole06}).

The comparable strength of the bow and upstream shocks in Abell 2146
suggests that this mass ratio could be lower than 4:1.  Closer
to an equal mass ratio is unlikely because the predicted peak Mach
number at core passage will then drop below the observed $M\sim2$.  We
therefore argue that the subcluster and main cluster components of
Abell 2146 are likely to have a mass ratio of 3 or 4:1.  New Subaru
Suprime-Cam observations of this cluster (PI Gandhi) will allow a more
quantitative analysis of the mass.

% Discuss similarities
% Why are we seeing this at same evolution time as bullet cluster?
% Would we see this object if it wasn't a merger?

\subsection{Subcluster structure}
% Bullet morphology
As the subcluster passed through the core of the main cluster, its
leading edge was compressed and swept back as the strong gravitational
potential stripped away the gas.  The local gas was pushed aside
during core passage producing an elongation of the gas distribution
perpendicular to the merger axis.  For high mass ratio mergers, a ring
of compressed gas forms when the small subcluster passes through the
primary core (\citealt{Roettiger97}; \citealt{Poole06}).  This can be
clearly seen as a bar-like structure in the observations of the Bullet
cluster (\citealt{Markevitch06}).  Figs. \ref{fig:labelled} and
\ref{fig:jssmaps} show an extended spur of cool gas at $5-7\keV$ to
the SW of the subcluster tail.  This is likely to be stripped material
from the subcluster and swept up gas from the main cluster.  However,
there does not appear to be a symmetric feature to the NE of the
subcluster tail, although there is a suggestion of a smaller spur to
the NE in Fig. \ref{fig:labelled}.  Simulations suggest that elongation
perpendicular to the merger axis may be less evident in lower mass
ratio mergers where the cores are similar sizes
(\citealt{Roettiger97}; \citealt{Poole06}).  A slightly off axis
merger will also produce an asymmetrical distribution of swept up and
stripped gas (\citealt{Poole06}).

% Mach cone angle
The bow shock visible in front of the subcluster core formed when the
subcluster's infall velocity exceeded the sound speed in the ambient
cluster gas, $c_s=1000^{+100}_{-90}\kmps$.  In principle, the Mach
cone angle should be directly related to the Mach number of the shock.
However, as has been found for Bullet cluster (\citealt{Markevitch02};
\citealt{Markevitch07}), the subcluster is shrinking over time and
decelerating in the gravitational potential of the main cluster.
Therefore, in practice the subcluster cannot be approximated as a
solid body moving at constant velocity.  \citet{Springel07} also found in their
simulations of the Bullet cluster that the opening angle of the Mach
cone was much wider than expected and cannot be easily used to
independently determine the shock strength.  For Abell 2146, we can only
estimate the Mach cone angle from the narrow section of the bow shock
that is easily discernable but it is clear that it is much broader
than the $\phi=30^{\circ}$ expected for a $M\sim2$ shock
(Fig. \ref{fig:mainimage}).

% Standoff distance, compare with bullet cluster
% Bullet cluster ~100kpc between bow shock and edge of bullet
We might also expect that the stand-off distance between the bow shock
and the subcluster is related to the Mach number, according to the
approximate relation of \citet{Moeckel49}\footnote{Available at
  http://naca.central.cranfield.ac.uk/reports/1949/naca-tn-1921.pdf}
(see also \citealt{Vikhlinin01}).  This analysis requires measurements
of the geometry of the projectile, which was particularly difficult to
determine for the subcluster gas cloud.  We estimated the half-width
of the subcluster core to be $\sim35\kpc$, which for $M=2.2\pm0.8$
gives a predicted stand-off distance between the shock and the
subcluster of $d\sim20-40\kpc$.  The observed distance is
approximately $140\kpc$.  A similar result can be obtained for the
Bullet cluster.  The assumptions of a uniform preshock medium,
constant velocities and a solid projectile are clearly not applicable
for these merging subclusters.

% Tail shocks (rarefaction + expansion)
Behind the subcluster, the ambient cluster gas that was pushed aside
during its passage will fall back and produce tail shocks
(eg. \citealt{Roettiger97}; \citealt{Poole06}).
Fig. \ref{fig:NWsectordeproj} shows a drop in density by a factor of
$1.91\pm0.07$ at the approximate position of the tail end
($\sim50\asec$ NW from the AGN).  If this feature were a shock we would
expect a corresponding jump in the temperature by a factor of
$1.66\pm0.07$.  Despite the uncertainties in the temperature values
(section \ref{sec:NWsector}) such a large decrease appears unlikely.
Alternatively, this feature could be a cold front separating the
cooler, ram pressure stripped material from the subcluster which is
slowly falling back into the hotter main cluster gas.  Although there
is a decrease in the electron pressure across the front, the large
uncertainties in the temperature value reduce the significance of this
drop.  A longer \emph{Chandra} observation providing a more accurate
measurement of the temperature could resolve this issue and allow
constraints on the ICM transport processes across the shocks and cold
fronts (eg. \citealt{Markevitch06}; \citealt{Markevitch07}).  

\begin{figure}
\centering
\includegraphics[width=0.98\columnwidth]{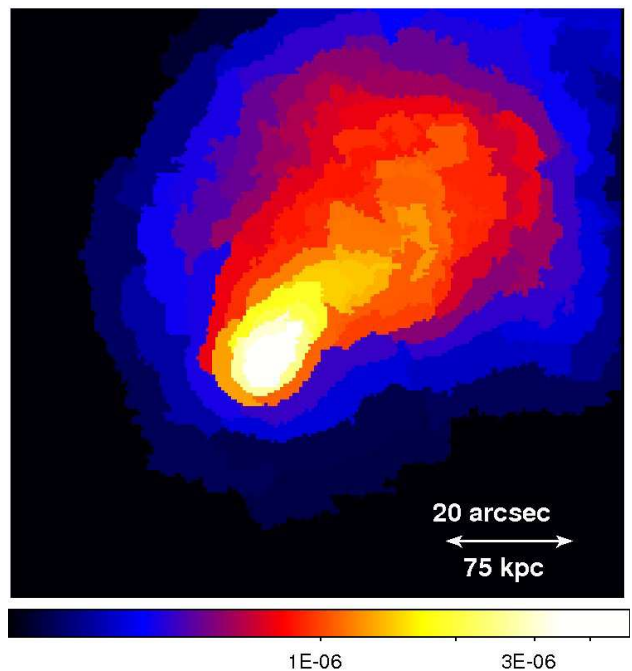}
\caption{Emission measure per unit area map (units
  $\empasecsq$).  The emission measure is the \textsc{xspec}
  normalization of the \textsc{mekal} spectrum
  $K=EI/(4\times10^{14}\pi D_A^2(1+z)^2)$, where EI is the emission
  integral $EI=\int n_en_H\mathrm{d}V$.}
\label{fig:normmapextra}
\end{figure}

% Instabilities
The subcluster will also generate strong turbulence in its wake.
Figs. \ref{fig:mainimage} and \ref{fig:zoomimage} show complex
structures in the ram pressure stripped material of the subcluster
tail.  The temperature map also suggests there could be substructures
and smaller shocks inside the tail.  The ram pressure stripped tail is
considerably warmer ($5-8\keV$) than the subcluster core where it
originated ($1.9\pm0.1\keV$).  This level of heating appears to be
consistent with the results for the Bullet cluster where the
subcluster core gas at $\sim2\keV$ is stripped and forms a tail of
gas heated to $7-10\keV$ (\citealt{Markevitch02};
\citealt{Million09}).  Along the surface of the bullet,
Kelvin-Helmholtz and Rayleigh-Taylor instabilities are expected to
develop and break up the subcluster, as has been observed in Abell 520
(\citealt{Markevitch05}).  The timescale for the destruction of the
subcluster will also likely be influenced by magnetic fields which
stabilize against these instabilities (eg. \citealt{Jones96};
\citealt{Vikhlinin01}).  Fig. \ref{fig:normmapextra} shows an emission
measure map of the cluster produced with the Contour binning algorithm
(section \ref{sec:contourbin}) but with a lower signal-to-noise ratio
of 15 giving better spatial resolution at the expense of larger errors
($\sim15\%$).  The absorption parameter was fixed to the Galactic
value and the abundance was fixed to $0.4\Zsun$ for the spectral
fitting.  The finer spatial binning in Fig. \ref{fig:normmapextra}
hints at possible substructure in the subcluster tail but a deeper
\emph{Chandra} observation is required to significantly detect any
structure.

% Do simulations show an ordered shock?

% Angle on sky, width of edges
% Impact parameter of collision
% Is the subcluster falling back?

\subsection{Location of the brightest cluster galaxy}
% Location of BCG
% Include image of DSS contours - discuss weak lensing possibility
In a merger event, the galaxies are effectively collisionless
particles and therefore, along with the dark matter component, lead
the X-ray gas on exiting the main cluster core.  The X-ray gas is
slowed as a result of the ram pressure from the interacting cluster
cores.  This is clearly observed in the Bullet cluster
(eg. \citealt{Clowe06}).  However, Fig. \ref{fig:zoomimage} shows that
while there are many galaxies leading the X-ray subcluster core in
Abell 2146, the BCG is located immediately behind the X-ray peak.

% Discuss ram pressure slingshot (discard because timescales wrong)
Observations of the merging cluster Abell 168 also found that the
subcluster cD galaxy lagged behind the cool gas peak
(\citealt{Hallman04}).  As the subcluster passes through the core of
the primary, ram pressure pushes back the gas from the gravitational
potential.  Then as the subcluster enters the outer, less dense
cluster gas, the ram pressure drops rapidly and the cool gas core
rebounds and overshoots the subcluster dark matter peak in a `ram
pressure slingshot' (\citealt{Hallman04}; \citealt{Mathis05};
\citealt{AscasibarMarkevitch06}).  The cD galaxy, which should trace
the local gravitational potential, will then appear behind the
subcluster's cool core.  However, this represents a late stage of the
merger where the subcluster has reached its apocentre.  For Abell
2146, this would occur $\sim1\Gyr$ (based on a mass ratio of 3:1,
\citealt{Poole06}) after the subcluster passes through the main
cluster core and is therefore inconsistent with the estimated
$0.1-0.3\Gyr$ for the age of the merger.  Although it was difficult to
determine the location of the main cluster core from the existing
observations, the prominence of the two shocks and the undisrupted
subcluster core provides strong evidence against this being a late
stage merger.  Unambiguous detections of shock fronts are so rare
because they can only be seen in the early stages of a merger, before
the shock front has propagated to the low surface brightness outskirts
of the cluster (\citealt{Markevitch07}).

% So large + extended that experiences drag?  As large as core.
The HST observation of Abell 2146 (contours shown in
Fig. \ref{fig:zoomimage}; \citealt{Sand05}) shows that the BCG
has a large halo of emission, which is extended along the merger axis
and clearly traces the X-ray emission in the subcluster.  The large
diffuse envelope of the BCG may therefore be interacting with the
surrounding X-ray gas.  The BCG does not obviously appear to be
disrupted by the merger event so it remains unclear why it is located in the
wake of the subcluster cool core.

% SFR comment
In an imaging survey with the \emph{Spitzer Space Telescope} of 62
BCGs with optical line emission located in the cores of X-ray luminous
clusters, \citet{Quillen08} found that Abell 2146 has a high IR
luminosity and the second highest rate of star formation in the sample
at $192\Msunpyr$.  \citet{ODea08} found a correlation between the mass
deposition rates estimated from X-ray observations of the sample and
the IR star formation rates, where the star formation rate is
$\sim1/10$ of the mass deposition rate.  This suggests that the
cooling ICM is the source of the gas that is forming stars (see also
\citealt{Johnstone87}; \citealt{McNamara04};
\citealt{HicksMushotzky05}; \citealt{Rafferty06}; \citealt{Salome06}).
However, in Abell 2146, the high star formation rate of $192\Msunpyr$
is $\sim5$ times greater than our estimated mass deposition rate of
$40\pm10\Msunpyr$.  This high rate of star formation in the BCG could have
been triggered by the cluster merger and this will be explored in a
future paper.

\subsection{Limits on diffuse radio emission}
The connection between cluster mergers and the presence of diffuse,
steep spectrum radio sources has been extensively investigated
(\citealt{Tribble93}; \citealt{Roettiger99}; \citealt{Brunetti09}) so
we have searched the radio survey data around Abell 2146 to determine
a limit to any diffuse emission. Unfortunately, Abell 2146 is just
below the X-ray luminosity limit set by \citealt{Venturi07} for their
$610\MHz$ GMRT survey so no targetted radio imaging at frequencies
below $5\GHz$ exists. However, if we take the observed upper bound for
radio halo power from \citealt{Brunetti09} for a cluster with the
X-ray luminosity of Abell 2146 ($P_{1.4\GHz}\sim 10^{24.0}\WpHz$) we
would expect a halo or relic that is weaker than $\sim 6\mJy$ at
$1.4\GHz$.  

Inspecting the VLSS $74\MHz$, WENSS $327\MHz$ and NVSS
$1.4\GHz$ survey data and archival VLA databases, we find two
unresolved sources detected in WENSS, NVSS and $5\GHz$ VLA images.
One is coincident with the BCG and is detected at $15.3\pm0.6\mJy$ at
$1.4\GHz$ with a spectral index of -0.43. The other is associated with
a probable cluster member that is $40.6\pm1.3\mJy$ at $1.4\GHz$ with a
spectral index of -0.48.  These two sources are blended in WENSS
catalog but the total flux density is consistent with the sum of the
two point source components.  Inspecting the noise in these maps we
calculate $3\sigma$ flux density limits of $<660\mJy$, $<21\mJy$ and
$<2.5\mJy$ at 74, 327 and $1420\MHz$ respectively for any diffuse
emission on scales of $1-2.5\amin$.  Therefore, a radio halo a factor
of around two below that expected from other comparable systems is
consistent with the current observations irrespective of the spectral
index. A deep eVLA or GMRT observation is required to improve on this
limit.

\section{Conclusions}
% Results
The \emph{Chandra} observation of Abell 2146 has revealed a merging
system where a ram pressure stripped subcluster has recently passed through
and disrupted the primary cluster core.  From the X-ray temperature
and surface brightness maps, we found a bow shock propagating in front
of the cool $2-3\keV$ subcluster and calculated a Mach number
$M=2.2\pm0.8$ from the density jump across the shock.  The subcluster
velocity is therefore $v=2200^{+1000}_{-900}\kmps$ and we estimated
that the subcluster passed through the main cluster core $0.1-0.3\Gyr$
ago.  In addition, there is a factor of $10^{+3}_{-2}$ drop in the
electron pressure in the outskirts of the main cluster indicating the
presence of a $M=1.7\pm0.3$ upstream shock.  There are
potentially further shocks and a cold front in the gas tail behind the
subcluster but these features could not be confirmed with the existing data.

% Formation scenario
Although Abell 2146 is a smaller and cooler system than the Bullet
cluster, it appears similar in structure and to be at a comparable
merger epoch of $\sim0.2\Gyr$ since core passage.  This can be
understood as a selection effect: shock fronts can only be detected at
this early stage in the merger evolution before they have propagated
to the outer, low surface brightness regions of the cluster.  The
merger axis must also be close to the plane of the sky so that
projection effects do not conceal the surface brightness edges.  

% In addition, by preferentially selecting the brightest X-ray clusters to
% observe, we are more likely to find clusters in this narrow window.
% Merging clusters undergo a dramatic increase in luminosity and
% temperature when their cores are compressed in the collision
% (\citealt{Ricker01}; \citealt{Ritchie02}).  Boosts in luminosity by up
% to a factor of 10 for a head-on collision between equal mass clusters
% are possible.  However, this is a short-lived phenomenon and decays
% after about one sound crossing time, which for a typical cool-core
% cluster is only $0.1\Gyr$.

% Outstanding problems + further work
Based on the measured shock Mach numbers of $M\sim2$ and the strength
of the upstream shock, we estimate a mass ratio between the two merging
clusters of around 3 or 4:1.  Forthcoming Subaru observations of Abell
2146 will allow a more quantitative analysis of the mass
distribution between the two clusters.  We compared the \emph{Chandra}
observation with an archival HST observation and found that while
there is a group of galaxies located in front of the X-ray subcluster
core, the brightest cluster galaxy is located immediately behind the
X-ray peak.  A future weak lensing analysis coupled with galaxy
velocities along the line of sight from optical spectroscopy
could help to explain the galaxy cluster dynamics.

\section*{Acknowledgements}
We acknowledge support from the Science and Technology Facilities
Council (HRR) and the Royal Society (ACF).  Support for this work was
provided by the National Aeronautics and Space Administration through
Chandra Award Number GO0-11012B issued by the Chandra X-ray
Observatory Center, which is operated by the Smithsonian Astrophysical
Observatory for and on behalf of the National Aeronautics Space
Administration under contract NAS8-03060.  We thank the referee for
helpful comments.

\bibliographystyle{mnras}
\bibliography{refs.bib}

\clearpage

\end{document}